\documentclass[acmsmall,authorversion]{acmart}
\acmJournal{PACMPL}
\acmVolume{}
\acmNumber{OOPSLA1}
\acmArticle{}
\acmYear{2024}
\acmMonth{1}
\acmDOI{} 
\startPage{1}

\setcopyright{rightsretained}

\usepackage{prelude}

\begin{document}

\title{Profiling Programming Language Learning}


\author{Will Crichton}
\orcid{0000-0001-8639-6541}
\author{Shriram Krishnamurthi}
\orcid{0000-0001-5184-1975}
\affiliation{
  \department{Department of Computer Science}
  \institution{Brown University}           
  \city{Providence}
  \state{Rhode Island}
  \postcode{02912}
  \country{USA}                    
}
\email{wcrichto@brown.edu}

\begin{abstract}
This paper documents a year-long experiment to ``profile'' the process of learning a programming language: gathering data to understand what makes a language hard to learn, and using that data to improve the learning process. We added interactive quizzes to \textit{The Rust Programming Language}, the official textbook for learning Rust. Over \data{13 months}, \data{62,526} readers answered questions \data{1,140,202} times. First, we analyze the trajectories of readers. We find that many readers drop-out of the book early when faced with difficult language concepts like Rust's ownership types. Second, we use classical test theory and item response theory to analyze the characteristics of quiz questions. We find that better questions are more conceptual in nature, such as asking why a program does not compile vs.\ whether a program compiles. Third, we performed \data{12} interventions into the book to help readers with difficult questions. We find that on average, interventions improved quiz scores on the targeted questions by \data{+20\%}. Fourth, we show that our technique can likely generalize to languages with smaller user bases by simulating our statistical inferences on small $N$. These results demonstrate that quizzes are a simple and useful technique for understanding language learning at all scales.
\end{abstract}



\maketitle

\section{Introduction}

Teaching prospective users is an inescapable part of programming language adoption. 
Yet, teaching PLs is more art than science.
PL learning resources are designed based on the intuitions of their authors. Feedback on these resources only comes at the macro-scale, such as whether programmers end up successfully adopting a language. This disconnect is becoming more salient as the learning curves for modern languages grow ever steeper. For instance, user surveys within the communities of OCaml\,\cite{ocamlsurvey2022}, Haskell\,\cite{haskellsurvey2022}, Rust\,\cite{rustsurvey2020}, Scala\,\cite{scalasurvey2022}, Clojure\,\cite{clojuresurvey2022}, and even Go\,\cite{gosurvey2023} all report that the language's learning curve is one of the biggest problems in the language's ecosystem.


Heeding the call of \citet{meyerovich2013adopt} to build a scientific foundation for ``socio-PLT,'' we set out to gather data about how people learn a new programming language, and to develop a generalizable methodology for improving PL learning resources. This paper reports on a year-long experiment to profile the process of PL learning within an online textbook. We use ``profile'' in the same sense as performance profiling for software --- our goal was to gather fine-grained data to help identify ``hot-spots'' where learners are struggling. Concretely, we studied \textit{The Rust Programming Language} (\trpl{}) \,\cite{trpl}, the official textbook for learning Rust. We chose to study Rust both because the Rust community has been exceptionally open to facilitating educational research, and because many people were seeking to learn Rust in 2022-23.

The central idea of the experiment is to add interactive quizzes to each chapter of \trpl{}, a total of \data{221} questions. The quizzes acted as profiling probes that gathered data about individual challenges faced by learners. From \data{September 2022} to \data{October 2023}, we gathered \data{1,140,202} answers from \data{62,526}+ participants. The contribution of this paper is analyzing this data to answer four questions:

\begin{enumerate}[leftmargin=0pt,itemsep=0.5em]
    \item[] \textbf{RQ1. What kinds of trajectories do readers take through the book?} (\Cref{sec:reader-analysis}) \\
    We find that the vast majority of readers do not reach the end of the book, consistent with data from MOOCs. We also find that difficult language concepts in early chapters (specifically, ownership in the case of Rust) serve as a common drop-out point for many readers.
    
    \item[] \textbf{RQ2. What are the characteristics of a high-quality PL quiz question?} (Sections \ref{sec:frequentist-question-analysis} and \ref{sec:bayesian-question-analysis}) \\
    We use both classical test theory (\Cref{sec:frequentist-question-analysis}) and item response theory (\Cref{sec:bayesian-question-analysis}) to model the difficulty and discrimination of each question.
    We find that the most discriminative questions focus on conceptual understanding over syntax or rote rules. In particular, questions about discerning well-typed vs.\ ill-typed programs were often \emph{not} discriminative.
    
    \item[] \textbf{RQ3. How can a learning profile be used to improve a PL learning resource?} (\Cref{sec:interventions}) \\
    We used \trpl{}'s learning profile to identify difficult questions, and then we created \data{12} interventions (small edits to the book) to help readers based on our theory of their misconceptions. We evaluated these interventions by comparing question scores before and after each intervention, finding that \data{10/12} interventions had a statistically significant effect with an average improvement of \data{+20\%}.

    \item[] \textbf{RQ4. How applicable is this methodology to languages with smaller user bases?} (\Cref{sec:simulations}) \\
    We test whether the quizzing methodology could work with languages that have user bases smaller than Rust's. We use random sampling and power analysis to simulate the above analyses for smaller $N$. We find that estimating reader drop-off, question characteristics, and intervention efficacy have relatively low error around $N = 100$, while estimating question discrimination requires larger $N$.
\end{enumerate}

\vspace{0.5em}

\noindent We first describe our experiment design (\Cref{sec:methodology}), and then analyze each RQ. We discuss threats to validity (\Cref{sec:threats}), related work (\Cref{sec:relatedwork}), and implications for  future research (\Cref{sec:discussion}).

\section{Experiment Design}
\label{sec:methodology}

To begin, we will describe the general setup of the learning platform used in the experiment.

\subsection{Design Goals}
\label{sec:design-goals}

Our general goal in the experiment was to identify patterns in how people learn a new programming language, and to use these patterns to improve the learning process. This general goal was guided by three design goals. These design goals are conflicting in several ways, so our experiment represents one point in a broader trade-off space of experimental designs. The goals are:

\begin{itemize}
    \item \textbf{Richness of data:} the experiment should generate data that provides a depth of understanding about the learning process, so as to improve the quality of the insights. For example, we consider that only tracking the time a learner spends on a given task would be insufficiently rich data.
    
    \item \textbf{Scale of participation:} the experiment should include as many participants as possible from a wide range of backgrounds, so as to make the results more statistically robust and more likely to reflect general trends than niche patterns. For example, we consider that analyzing a classroom of students at one university would be an insufficiently large scale of participation.
    
    \item \textbf{Simplicity of infrastructure:} the setup for the experiment should require as little cost and complexity as possible, to help authors of other learning resources replicate our methodology. For example, methods that require expensive servers or complex backends would be insufficiently simple. 
\end{itemize}

One immediate consequence of these goals is that they rule out methods which require compensation for participants. Compensation is neither scalable on an academic budget, nor is it logistically simple to disburse to a global audience. Instead, we sought to find an experiment design which would intrinsically motivate people to participate.

\subsection{Selecting a Learning Environment}

The foundation for the experiment is a programming language \emph{learning environment}, or a place where language learning regularly occurs such that we can study the process. To maximize scale of participation, we looked for widely-used web-based learning environments. As of 2023, surveys suggest that \emph{online textbooks} are the most popular long-form learning resource for developers\,\cite{sosurvey2023,jbsurvey2022}, so it is a natural environment for study. We chose to study \emph{The Rust Programming Language} (\trpl{})\,\cite{trpl}, the official textbook for learning Rust. Rust is an up-and-coming language for safe systems programming, and Rust is an especially good candidate for study because its learning curve has been consistently reported as a major barrier for adoption\,\cite{rustsurvey2020,fulton2021,zhu2022,zeng2018}.

\trpl{} consists of 20 chapters that cover Rust from the basics (variables, control flow) to intermediate topics (ownership, traits) to advanced topics (macros, unsafe code). \trpl{} is oriented at readers with some programming experience --- the book ``assumes that you’ve written code in another programming language but doesn’t make any assumptions about which one''\,\citeyearpar[Introduction]{trpl}. 

\trpl{} is an open-source textbook written in Markdown using the mdBook framework. mdBook converts Markdown chapters into HTML files linked by a shared table of contents, similar to tools such as Sphinx and Scribble. mdBook is also used by other textbooks such as \textit{Theorem Proving in Lean 4}\,\cite{tpil4} and \textit{Comprehensive Rust}\,\cite{comprehensiverust}.
We forked the GitHub repository for \trpl{} on \data{October 22, 2022} and setup a web page hosting our forked version at the URL: \url{https://rust-book.cs.brown.edu/}. Because mdBook does not require a backend service (unlike e.g., a MOOC platform), we only needed to find a CDN to distribute the static files generated by mdBook. We chose to use GitHub Pages because it is both robust and free of charge as of 2023.

\subsection{Adding Quiz Questions}

The baseline \trpl{} environment does not provide any data on a reader's learning process, so we needed to enrich the environment with learning probes. We decided to add frequent, short quizzes throughout the textbook. \Cref{fig:example-question-review} shows an example quiz question as it appears in our \trpl{} fork. Quiz questions satisfy each of our design goals:
questions provide rich data through content-tailored probes of a reader's knowledge.  Questions incentivize participation for readers who seek an interactive learning experience. And questions are simple to write and deploy at scale.

We designed both the quiz questions and the quizzing tool from scratch for this experiment. Of course, our quiz questions are similar to what any experienced educator might think to create, and our tool is in part just a reimplementation of a standard quizzing interface. But we still want to highlight some key ways in which the questions and the tool relate to our design goals.

\subsubsection{Quiz questions}
\label{sec:quiz-questions}

\begin{figure}
    \centering
\begin{subfigure}{0.43\textwidth}
        \begin{qminted}{toml}
[[questions]]
id = "1665d1ef-961f-4451-a988-ec46121531f9"
type = "MultipleChoice"
prompt.prompt = """
Which call to `find_until` function will
cause a runtime panic?
```
fn find_until(
  v: &Vec<i32>, n: i32, til: usize
) -> Option<usize> {
  for i in 0 .. til {
    if v[i] == n {
      return Some(i);
    }
  }
  return None;
}
```
"""
answer.answer = 
  "`find_until(&vec![1, 2, 3], 4, 4)`"
prompt.distractors = [
  "`find_until(&vec![1, 2, 3], 0, 0)`",
  "`find_until(&vec![1, 2, 3], 3, 3)`",
  "`find_until(&vec![1, 2, 3], 1, 4)`"
]
context = """
If `til = 4`, then for a vector of length 3, 
the for-loop will attempt to index the vector 
with `i = 3`, which is out of bounds. This 
function does not panic if `n = 1` because 
it returns before reaching the out-of-bounds 
index.
"""
    \end{qminted}
    \cprotect\caption{The TOML format for authoring questions. Metadata like \verb|id| is used for telemetry.}
    \label{fig:example-question-toml}
    \end{subfigure}
    \hfill
    \begin{subfigure}{0.53\textwidth}
        \includegraphics[width=\textwidth]{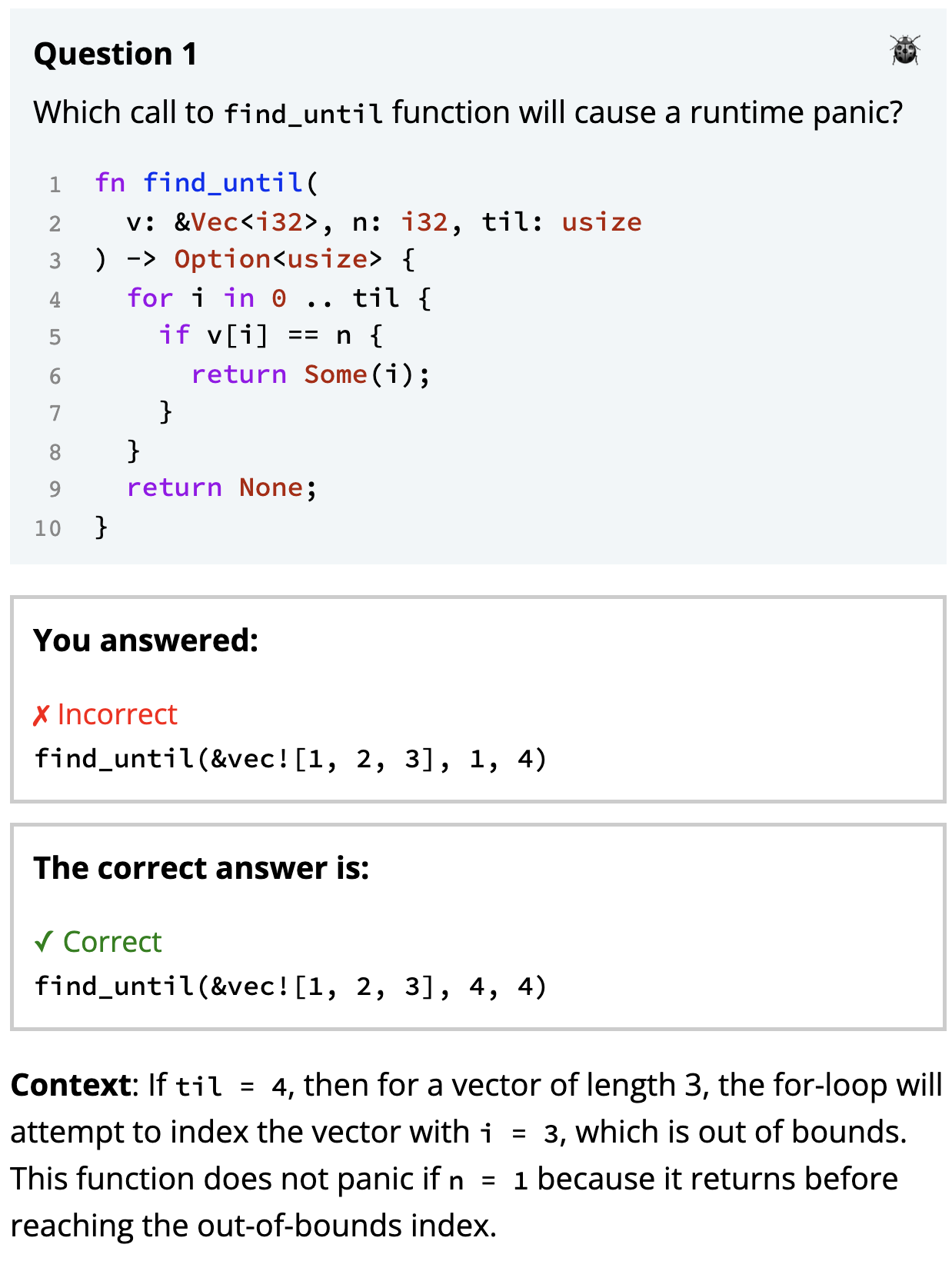}
        \caption{The answer review screen shown to readers after attempting to answer a question.}
        \label{fig:example-question-review}
    \end{subfigure}
    \subfigtocapspace{}
    \caption{An example question about vectors and bounds-checking. \data{83\%} of readers answer this question correctly.}
    \label{fig:example-question}
\end{figure}

The core of our experimental materials is the set of \data{88} quizzes consisting of \data{213} questions that we wrote for \trpl{}. We added at least one quiz to every section (i.e., standalone web page) of \trpl{}, except for sections that were extremely short (e.g., Section 3.4 ``Comments'') or sections that were intended as worked examples rather than introduction of new material (e.g., Section 2 ``Programming a Guessing Game''). The questions were evenly spread throughout the book, with an average of \data{2.4} questions per quiz. The first author wrote all questions, and we estimate that the authoring process took roughly 50 person-hours of effort.

The questions can be decomposed along three dimensions: format (the structure of a question), complexity (the number of steps in solving a question), and content (what a question is about).


\paragraph{Format} 

We chose to only use close-ended questions so readers could get immediate feedback on their answers, providing additional incentive to participate in the experiment. (Future work may consider either crowd-sourcing or machine learning for evaluation of open-ended responses, but that is outside the scope of the present work.) Specifically, we used three question types:

\begin{itemize}
    \item \textbf{Multiple choice:} A question with one correct answer and multiple incorrect answers, or \emph{distractors}. Readers select one answer from a list of all options. The order of options is randomized by default. \mdbookquiz{} also supports multiple-\textit{select} questions that have potentially multiple correct answers, and users must select the subset of options that are correct.

    \item \textbf{Short answer:} A question with an answer that is a precise string, and readers must respond with that string (usually modulo whitespace and case-sensitivity). In practice, we use the short answer type for either syntax questions (e.g., ``what is the keyword for declaring a function?'') or for numeric questions (e.g., ``how many allocations could occur in this program?'').

    \item \textbf{Tracing:} A question about the semantics of a particular program (in this experiment, Rust). Readers are given a program and asked: does this program compile\footnotemark{}? If so, what does it print to stdout? Like with short-answer, readers must provide the stdout string precisely modulo whitespace and casing. 
\end{itemize}

\footnotetext{
    Early in the experiment, tracing-type questions also asked: if the program does not compile, what line does the compiler report as the main source of the error? We ultimately removed this portion of the question because many readers did not understand the distinction between \textit{the point at which the error occurred} versus \textit{a line that is involved in the error}. The notion of a ``main source'' is also somewhat arbitrary in certain cases. Nonetheless, we still hope to eventually find a version of this question that elicits a next level of depth about \textit{why} a program failed to compile.
}

\paragraph{Complexity}
Our general philosophy was to test for what educational psychologists call \emph{near transfer}\,\cite{transfer1994}, or the application of knowledge to ``closely related contexts'' compared to the context of learning.
For the most part, we avoided questions that ask readers to simply repeat back a definition or formula contained verbatim in the text. Conversely, we also avoided questions that require paragraphs of setup with many moving parts. We sought a middle ground that asked readers to apply the book's concepts to slightly novel situations. For instance, the question in \Cref{fig:example-question} asks readers to apply the concepts of range-based iteration and bounds-checked indexing in a program that is not similar to any within the section on vectors.

\paragraph{Content}
Most of our questions can be categorized based on the taxonomy of programming language concepts in common use within PL research: syntax (concrete and abstract), semantics (static and dynamic), and pragmatics. We wrote questions to test whether readers understood a given language feature (e.g., algebraic data types, traits, etc.) at each level of the taxonomy.  For instance, these are a few common question archetypes:
\begin{itemize}
    \item \textbf{Concrete syntax}: ``what is the keyword for a given feature?''
    \item \textbf{Abstract syntax}: ``which binding does this variable refer to?''
    \item \textbf{Static semantics}: ``which of these programs will compile?'' or ``what kind of compiler error would you expect to get for this program?''
    \item \textbf{Dynamic semantics}: ``what is the output of running this program?'' or ``which of these programs will cause undefined behavior?'' or ``which lines of code can cause a heap allocation?''
    \item \textbf{Pragmatics}: ``which of these type signatures best matches this prose specification?'' or ``which of these changes to a broken function is the most idiomatic?''
\end{itemize}

\subsubsection{Quiz Tool}
\label{sec:quiz-tool}

To integrate into \trpl{}, we developed \mdbookquiz{}, a plugin that extends mdBook with support for interactive quizzes. To use \mdbookquiz{}, an author first writes quiz questions as a TOML file like in \Cref{fig:example-question-toml}. Within their mdBook, the author then adds a directive like {\small\verb|{{#quiz ../quizzes/my-quiz.toml}}|}. The \mdbookquiz{} preprocessor replaces the directive with an HTML element containing the quiz schema as metadata. When a reader loads a page containing a quiz, then \mdbookquiz{} injects an interactive quiz at each such HTML element.
The quiz component is implemented in Javascript using React. The notable elements of its design are:

\paragraph{Session tracking}
It is important for rich data to track when a single reader is answering multiple questions, as that enables analysis of reader-level accuracy. However, we did not want to require readers to register accounts (disincentivizing participation), or to build the services needed to maintain account information (complex infrastructure). As a trade-off, we use cookies to associate a reader with a random UUID, stored in the browser's  {\small\verb|localStorage|}. This ID requires no input from users, but we can lose continuity if a reader switches devices/browsers or clears their cookies.

\paragraph{Telemetry} 
When a reader completes a quiz, the quiz sends telemetry to a server. The telemetry includes the reader's answers, a timestamp, the commit hash of the \trpl{} fork (to know which version of the book an answer corresponds to), and the reader's ID. For the remote server, we setup a small Google Cloud VM that streams the telemetry to a SQLite database. This server is the only component of the experiment that cost us money, a relatively small \$15/month at 2022-23 rates.

\paragraph{Mobile support}
To maximize participation, we must support the popular modes for web browsing. Many \trpl{} readers visit the book on their phones, so we took strides in \mdbookquiz{} to support non-desktop devices. \mdbookquiz{} uses responsive styles to ensure that the quizzes remain legible at all resolutions. We also ensure that interactions work with both touch and click inputs.

\paragraph{Question validation} 
Faulty questions lead to faulty data, so it is important to correct mistakes as soon as possible. We used two mechanisms to detect bad questions:
\begin{itemize}
    \item \textbf{Schema checking:} the \mdbookquiz{} tool checks at book-generation-time that all questions adhere to the expected schema. This includes semantic checks for tracing-type questions, where the expected answer (i.e. compiles with output OR fails to compile) is compared against the Rust compiler's actual results.
    \item \textbf{Bug reporting:} Every question is annotated with a bug button as shown in the upper right of \Cref{fig:example-question-review}. Readers can provide open-ended feedback, which we have found valuable in catching issues such as typos, confusing problem wording, and faulty questions. To date, readers have made \data{2,026} reports. Throughout the experiment, we continuously monitored these reports and patched questions when needed.
\end{itemize}

\paragraph{Reducing external influence} 
A challenge with a public-facing textbook is the quizzing environment is largely uncontrolled, so participants could easily use external resources (the textbook, Google, ChatGPT, etc.). To combat this threat to data richness, we took two steps. First, when a reader begins a quiz, the quiz component takes over the entire page, preventing readers from easily using the textbook while answering questions. Second, we instruct readers to not use external resources while answering questions. While this relies on voluntary compliance, our readers also have little incentive to cheat considering the quizzes are solely for personal edification.

\paragraph{Retries} 
After giving an initial answer to all the questions in a quiz, readers are offered the chance to retry missed questions. For the analysis in this paper, we only look at answers from readers' first attempt on a quiz. But it worth noting that when offered the option, readers chose to retry a quiz in \data{55\%} of cases, suggesting that readers appreciate the opportunity to correct their mistakes.

\paragraph{Answer justification}
To provide richer data for particular questions, authors can optionally enable an ``answer justification mode'' for a given question. In this mode, after answering a question and before seeing the correct answer, a reader will be prompted to explain how they chose their answer in 1-2 sentences. This justification is then sent with the telemetry.

\subsection{Recruiting Participants}

Once the quiz questions were deployed to our \trpl{} fork, the final step was to recruit participants. Our primary recruitment channel was the official \trpl{}, where the \trpl{} authors graciously agreed to include the following ad: ``Want a more interactive learning experience? Try out a different version of the Rust Book, featuring: quizzes, highlighting, visualizations, and more: \url{https://rust-book.cs.brown.edu/}''

On \data{August 25, 2022}, we soft-launched the experiment by announcing the URL on the first author's Twitter account. On \data{November 1, 2022}, the advertisement for our experiment went live in the official web version of \trpl{}. The advertisement remained there throughout the experiment and continues to drive traffic as of \data{October 9, 2023}. Since the ad went live, our website has received at an average of \data{439} visitors per day.

Again following from the goal of increasing scale of participation, we did not gather any demographic data from our participants. Our dataset is fully anonymized except for the UUID that accompanies all telemetry. This has the benefit of encouraing privacy-conscious readers to participate in the experiment, but has the drawback of reducing data richness because we cannot slice the data based on demographic factors like a reader's level of programming experience. 


When prospective readers first visit our \trpl{} fork, they are provided a consent form which explains the purpose of the experiment and the kind of data collected from their interactions. Readers must provide consent in order to proceed to the book proper.  Our methodology was evaluated by our institution's IRB, which determined that the project did not require institutional review due to the study's purpose and safeguards to ensure anonymity.




\section{Data Analysis}
\label{sec:data-analysis}

Using the \trpl{} fork described in \Cref{sec:methodology}, we gathered reader responses to quiz questions for \data{13 months}. In total, \data{62,526} people answered questions \data{1,140,202} times. In this section, we will analyze the reader-level and question-level trends in the data to answer our first two research questions.

\subsection{Reader Analysis}
\label{sec:reader-analysis}

\begin{figure}
    \centering
    \begin{minipage}[t]{0.43\textwidth}
        \includegraphics[width=\textwidth]{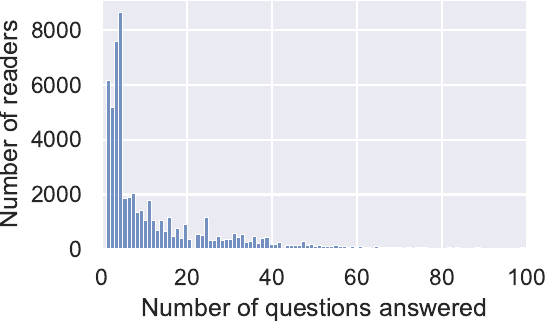}
        \captionof{figure}{Histogram of the number of questions answered per reader.}
        \label{fig:answers-by-user}
    \end{minipage}
    \hfill
    \begin{minipage}[t]{0.54\textwidth}
        \includegraphics[width=\textwidth]{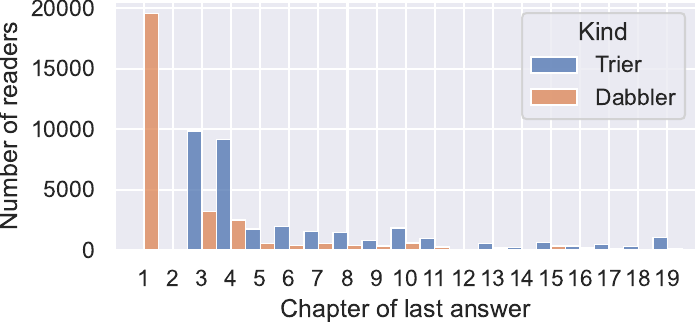}        
        \captionof{figure}{Histogram of chapter of a reader's last answered, broken down by triers vs. dabblers.}
        \label{fig:last-chapter}
    \end{minipage}
    \vspace{-1em}
\end{figure}

First, we will get to know the readers of \trpl{} in answering RQ1: what kinds of trajectories do readers take through the book?  We will start by looking at the amount of effort put in by each reader. The best proxy for effort in our dataset is the number of questions answered.

\Cref{fig:answers-by-user} shows a histogram of the number of questions answered per reader, revealing two insights. \Cref{fig:answers-by-user} shows that many readers answer a small number of questions. As expected for a public resource, many people try learning briefly and then move on. We define a reader as a ``dabbler'' if they answered fewer than the median number of questions (\data{6}). Conversely, a reader is a ``trier'' if they answered 6 or more questions. By this definition, our dataset contains \data{33,013} triers. 

\Cref{fig:answers-by-user} shows that even the most voracious readers answered fewer than 100 questions, which is less than half of the total number of questions in the book. This observation could be explained in a number of ways: readers give up early, readers skip around in the book, or readers change browsers/devices and lose their session ID. To explore the first explanation further, we can look at the last chapter where a reader answered a quiz question.

\Cref{fig:last-chapter} shows a histogram of the number of readers whose last answered question was on a given chapter, broken down by triers vs. dabblers. One observation is that the vast majority of dabblers gave up on Chapter 1 while most triers gave up on Chapters 3 and 4. This data supports the dabbler vs.\ trier dichotomy as most dabblers gave up immediately while triers read further.

Another observation about \Cref{fig:last-chapter} is that even for triers, most did not make it past Chapter 4. For context, Chapter 4 is the chapter on ownership, which is Rust's approach to memory management. Ownership is fundamental to Rust's design, but prior work also suggests that ownership is one of the language's more difficult concepts to learn\,\cite{rustsurvey2020,fulton2021}. \Cref{fig:last-chapter} is consistent with this finding, as Chapter 4 is a significant drop-off point for triers. Interestingly, the drop-off does not change linearly --- once readers make it past Chapter 4, the distribution of drop-offs is roughly uniform over the remaining chapters.

Finally, \Cref{fig:last-chapter} shows that very few readers make it to the end of the book. Out of \data{62,526} readers, only \data{1,220} (\data{2\%}) answered a question in Chapter 19. Again, several possible explanations could apply. One is that most people simply do not want to put in the effort to read the entire book. The 2\% number is comparable to drop-out rates in MOOCs\,\cite{jordan2014mooc}, so this may reflect a ``natural'' effort distribution for online learning resources. Another explanation is that readers do not intend to read the book cover-to-cover. One could easily imagine a programmer reading enough of the book to write a nontrivial Rust program, and then only visiting later chapters as needed for reference. The data does not allow us to easily distinguish between these explanations, but it does point to a research question for later investigation.
\begin{takeaways}
    \item The vast majority of readers will only read the first few chapters and then give up. Focus on reducing learning obstacles in the first few chapters.
    \item Difficult language concepts that appear early on will serve as a natural drop-off point. Consider spreading out the difficulty, or perhaps motivating readers to continue reading even with a partial understanding of the difficult concept.
\end{takeaways}

\subsection{CTT Question Analysis}
\label{sec:frequentist-question-analysis}

Next, we will analyze RQ2: what are the characteristics of a high-quality PL quiz question? ``Quality'' is a naturally subjective term, but we can initially orient our analysis by looking at how researchers study quiz questions in other disciplines. The field of \emph{psychometrics} has long studied the quality of questions within contexts such as standardized tests and surveys\,\cite{raykov2011}. 

Within psychometric item analysis, there are three key concepts: an \emph{item} (i.e., a question), an \emph{instrument} (i.e., a set of items), and a \emph{latent trait}. The general idea is that the goal of an instrument is to measure this latent (or unobservable) trait in an individual. In educational settings, the latent trait is usually how well a person understands a particular concept or topic, often short-handed as ``ability.'' Within this framework, a question is often evaluated in terms of two properties:
\begin{itemize}
    \item \textbf{Difficulty:} the level of ability required to correctly answer a question.
    \item \textbf{Discrimination:} how well a question distinguishes between high and low ability readers.
\end{itemize}

For example, a question in \trpl{} like ``which number between 1 and 10 am I thinking of?'' would be high difficulty (only about 10\% of people should answer correctly) and low discrimination (correctly answering this question has nothing to do with Rust). Note that difficulty and discrimination are not mathematical quantities by definition, but rather theoretical constructs which are realized by specific mathematical models. In this paper, we will consider two models: \emph{classical test theory} (CTT) based on a frequentist analysis of descriptive statistics, and \emph{item response theory} (IRT) based on a Bayesian analysis of a probabilistic model.

In CTT, a reader's ability is the average of their raw scores on all questions. A question's difficulty is the average of reader's raw scores on the question. A question's discrimination is the correlation (Pearson's $r$) between reader's scores on that question and their overall score. The advantage of CTT is that the statistics are simple and interpretable, so we do not run the risk of ``torturing'' the data. The disadvantage of this approach is that the simplicity loses nuance, such as the idea that a reader correctly answering a difficult question should indicate a higher ability than correctly answering an easy question.

Note that in this section, we will only analyze responses from triers, as we care most about the learning patterns of readers who seriously attempted the book. We will use ``reader'' synonymously with ``trier'' in Sections \ref{sec:frequentist-question-analysis} and \ref{sec:bayesian-question-analysis}.

\subsubsection{CTT Ability and Difficulty}

\begin{figure}
    \centering
    \begin{minipage}[t]{0.47\textwidth}
        \includegraphics[width=\textwidth]{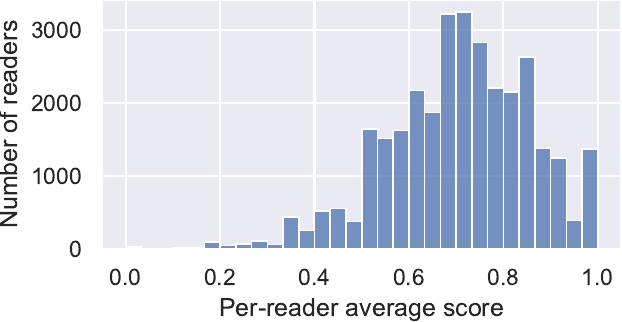}
        \captionof{figure}{Distribution of CTT reader ability.}
        \label{fig:reader-average-scores}
    \end{minipage}
    \hfill
    \begin{minipage}[t]{0.47\textwidth}
        \includegraphics[width=\textwidth]{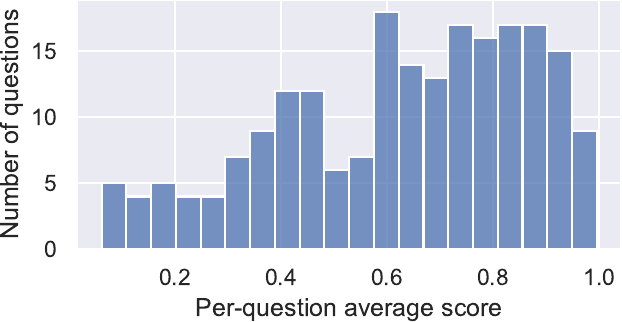}
        \captionof{figure}{Distribution of CTT question difficulty.}
        \label{fig:question-average-scores}
    \end{minipage} 
    \vspace{-1em}
\end{figure}

\Cref{fig:reader-average-scores} shows the distribution of reader ability and \Cref{fig:question-average-scores} shows the distribution of question difficulty using CTT metrics. Reader abilities are roughly normally distributed with mean \data{0.70} ($\sigma = \data{0.16}$). Question difficulties are more spread out with mean \data{0.63} ($\sigma = \data{0.24}$). The reader distribution in \Cref{fig:reader-average-scores} seems appropriate for the \trpl{} setting. Questions should be difficult enough to be engaging, but not so difficult as to potentially disincentivize readers from continuing; \Cref{fig:reader-average-scores} is consistent with that philosophy. The left tail of the question difficulty distribution is more concerning, as any question with less than a 20-30\% average indicates a likely problem with the question, with the text, or with the underlying concept.

\begin{figure}
    \centering
    \begin{subfigure}[t]{0.38\textwidth}
        \begin{question}
            \questiontype{Tracing}
            \begin{questionprompt}
                \begin{qminted}{rust}
fn foo(x: &i32) { 
  println!("{x}");
}

fn main() {
  let x = null;
  foo(x);
}
\end{qminted}
            \end{questionprompt}
            \begin{qchoices}
                \wronganswer DOES compile
                \rightanswer Does NOT compile
            \end{qchoices}
        \end{question}
        \caption{A low-difficulty question with an average accuracy of \data{99.5\%}.}
        \label{fig:least-difficult-question}
    \end{subfigure}
    \hfill
    \begin{subfigure}[t]{0.59\textwidth}
        \begin{question}
            \questiontype{Multiple Select}
            \begin{questionprompt}
                Which of the following are situations where using \rss{unsafe} code (or a safe wrapper around \rss{unsafe} code) is an idiomatic method for working around the borrow checker?
            \end{questionprompt}
            \begin{qparachoices}
                \rightanswer Getting two mutable references to disjoint indices in an array
                \rightanswer Allowing values to be uninitialized when they are not being read
                \rightanswer Having a reference to one field of a struct sit in another field of the same struct
                \wronganswer Returning a pointer to a stack-allocated variable out of a function
            \end{qparachoices}
        \end{question}
        \caption{A high-difficulty question with an average accuracy of \data{6.9\%}.}
        \label{fig:most-difficult-question}
    \end{subfigure}
    \subfigtocapspace{}
    \caption{Examples of low and high difficulty questions in the dataset.}
    \label{fig:difficult-questions}
\end{figure}

Placing these numbers in context, \Cref{fig:difficult-questions} shows examples of low and high difficulty questions in the dataset. The low difficulty question in \Cref{fig:least-difficult-question} tests whether a reader understands that (safe) Rust does not have null pointers, and \data{99.5\%} of readers correctly identify the program does not compile. This seems like a good result, but the near-absence of failure suggests that the question may be too easy due to meta-textual cues. For instance, the fact that \rs{null} does not get syntax-highlighted hints that it is not a language keyword. Additionally, it's not clear what the stdout would be if the program were executed, so a person may also infer that the question only makes sense if it does not compile. A better version of this question would ideally eliminate these cues.

The high difficulty question in \Cref{fig:most-difficult-question} tests whether a reader understands the purpose of the \rs{unsafe} feature in Rust. The answers includes two situations that are described in the textbook (mutable array indexing and returning pointers to the stack), and it asks readers to transfer their understanding to the other two novel situations (which are both reasonable cases to use unsafe). \data{94\%} of readers select the array option and \data{16\%} of readers select the stack-reference option, which is good. The failure mode is that only \data{25\%} and \data{34\%} of readers select the uninitialized data and self-referential struct options, respectively. In total, only \data{6.9\%} of readers select the correct subset of options --- quite close to the accuracy of random guessing (6.25\%). This suggests that the book can improve in helping readers transfer their understanding of \rs{unsafe} into these novel situations.

\subsubsection{CTT Discrimination}
\label{sec:ctt-discrimination}

\begin{figure}
    \centering
    \begin{minipage}[t]{0.47\textwidth}
            \includegraphics[width=\textwidth]{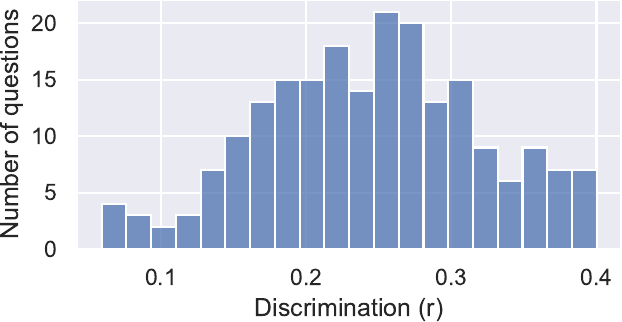}
        \captionof{figure}{Distribution of CTT question discrimination.}
        \label{fig:question-correlation}
    \end{minipage}
    \hfill
   \begin{minipage}[t]{0.47\textwidth}
        \includegraphics[width=\textwidth]{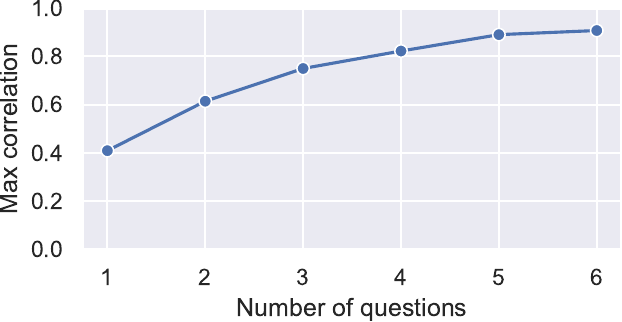}
        \captionof{figure}{Highest correlation of a subset of questions of size $k = 1$ to $6$.}
        \label{fig:test-construction}
    \end{minipage}
    \vspace{-1em}
\end{figure}

\Cref{fig:question-correlation} shows the distribution of question correlations. Questions range from $r = \data{0.06}$ to $r = \data{0.40}$. No questions had a negative correlation, which is a good sign. 
%
%
%
To better understand the characteristics of low/high discrimination questions, we performed a thematic analysis\,\cite{thematicanalysis} of the top 20 and bottom 20 questions by discrimination. We coded each question with labels that categorized the question such as ``undefined behavior'' and ``traits.'' We coded in multiple iterations, applying new labels and merging similar labels each time.

\begin{figure}
    \centering
    \begin{subfigure}[t]{0.35\textwidth}
        \begin{question}
            \questiontype{Tracing}
            \begin{questionprompt}
\begin{qminted}{rust}
fn main() {
  let v = vec![
    String::from("Hello ")
  ];
  let mut s = v[0];
  s.push_str("world");
  println!("{s}");
}
\end{qminted}
            \end{questionprompt}
             \begin{qchoices}
                \wronganswer DOES compile
                \rightanswer Does NOT compile
            \end{qchoices}
        \end{question}
        \caption{A low correlation type-error question with $r = 0.12$.}
        \label{fig:low-correlation-type-error}
    \end{subfigure}
    \hfill
    \begin{subfigure}[t]{0.63\textwidth}
        \begin{question}
            \questiontype{Multiple Choice}
            \begin{questionprompt}
\begin{qminted}{rust}
fn get_or_default(arg: &Option<String>) -> String {
  if arg.is_none() {
    return String::new();
  }
  let s = arg.unwrap();
  s.clone()
}
\end{qminted}

\noindent If you tried to compile this function, which of the following best describes the compiler error you would get?
            \end{questionprompt}
            \begin{qparachoices}
                \rightanswer cannot move out of \rss{arg} in \verb|arg.unwrap()|
                \wronganswer cannot call \verb|arg.is_none()| without dereferencing \rss{arg}
                \wronganswer \rss{arg} does not live long enough
                \wronganswer cannot return \verb|s.clone()| which does not live long enough
            \end{qparachoices}
        \end{question}
        \caption{A high correlation type-error question with $r = 0.37$.}
        \label{fig:high-correlation-type-error}
    \end{subfigure}
    \subfigtocapspace{}
    \caption{Examples of high and low correlation questions about type errors.}
    \label{fig:type-error-correlations}
    \vspace{-1em}
\end{figure}

One notable trend we observed is that \data{6/20} low-discrimination questions were tracing-type questions where the answer was ``does not compile'' due to a type error, while \data{0/20} high-discrimination questions used this format. \Cref{fig:type-error-correlations} contrasts two questions with low and high discrimination that both involve a type error. The low-correlation question posits that a program might compile, and a reader must deduce \emph{that} the program does not compile (because \rs{v[0]} moves the string out of the vector, which Rust disallows). The high-correlation question posits that a program does not compile, and a reader must deduce \emph{why} the program does not compile. We interpret this result as meaning: type-error questions that function as ``gotchas'' questions (you thought this would compile --- but it doesn't!) are significantly less useful in evaluating a reader's understanding of the type system than type-error questions that probe for deeper reasoning (why doesn't this compile?).

Another use of discrimination is in the design of instruments: what is the smallest set of questions that would make an effective test of Rust knowledge? For instance, one could use this data to construct a pre-test for a Rust course, or to construct a job interview for Rust professionals. To investigate, we ran an exhaustive combinatorial search for $k = 1$ to $6$ to find the $k$-sized subset with the highest correlation between user's average scores on the subset and their average scores on the entire dataset. ($k = 6$ is the highest we could compute in under a day on an M1 Macbook Pro.)
\Cref{fig:test-construction} shows the highest $r$ for each $k$. At $k = 6$ we can reach $r = 0.91$, showing that only a small number of questions is needed to achieve remarkably high discrimination. \Cref{sec:highest-correlating-subset} shows the six questions in this subset. This method of test construction is not foolproof given the issues with discrimination discussed in \Cref{sec:construct-validity}, but it offers one signal for designing tests from quiz data.

\begin{takeaways}
    \item The distributions of reader averages and question averages provide a useful signal to check whether outliers are too extreme, e.g., if some questions are too easy or if some readers are struggling too much.
    \item Asking \emph{why} a program has a given semantics is likely more discriminative than asking \emph{which} of several semantics a program could have.
\end{takeaways}

\subsection{IRT Question Analysis}
\label{sec:bayesian-question-analysis}

An alternative approach to modeling question difficulty and discrimination comes from the paradigm of \emph{item response theory} (IRT)\,\cite{irt2001}. The basic idea of IRT is that ability, difficulty, and discrimination can be incorporated into a single probabilistic model which is then fit to data. By fitting all parameters simultaneously, the resultant model can account for nuances which were missing with the CTT approach, such as the fact that multiple-select questions are harder to guess than multiple-choice questions due to the relative sizes of the combinatorial answer space.

\subsubsection{Background}

Formally, for some question $i$ and reader $j$, let $\mathbbm{1}_{ij}$ be an indicator random variable for whether $j$ answers $i$ correctly. Let $z_i$ be the properties of question $i$ and $\theta_j$ be the properties of reader $j$. IRT defines a family of functions $f(z, \theta)$ that describe the probability of $\mathbbm{1}_{ij}$:
$$
P[\mathbbm{1}_{ij}] = f(z_i, \theta_j)
$$

In this paper, we will use the standard \emph{three parameter logistic model} where $z_i = (\alpha_i, \beta_i, \lambda_i) : \mathbb{R}^3$ is a set of three parameters representing a question's discrimination, difficulty, and baseline, and $\theta_j : \mathbb{R}$ is a reader's real-valued ability score. The model's probability function is defined as:
$$
f((\alpha_i, \beta_i, \lambda_i), \theta_j) = \lambda_i + (1 - \lambda_i) \cdot \sigma(\alpha_i(\theta_j - \beta_i))
$$
Where $\sigma$ is the sigmoid function $\sigma(x) = \frac{1}{1 + e^{-x}}$.

\begin{figure}
    \centering
    \includegraphics[width=\textwidth]{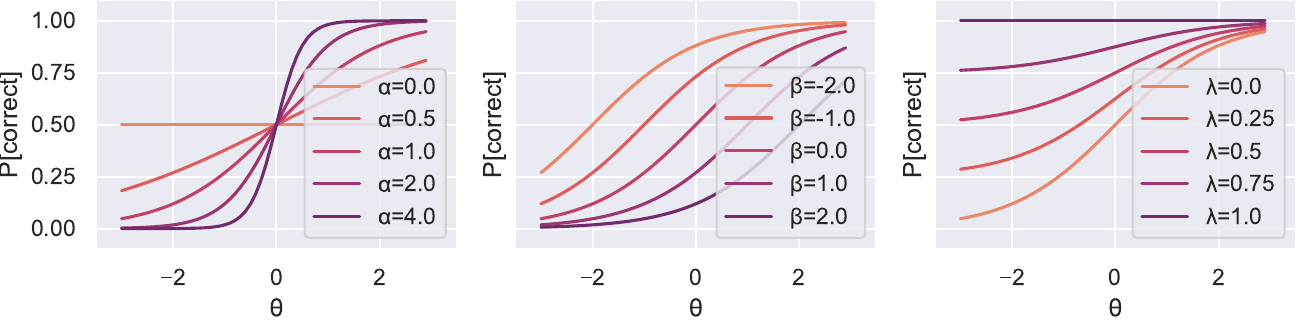}
    \vspace{-1em}
    \caption{Simulated item characteristic curves that demonstrate the influence of the discrimination parameter $\alpha$, the difficulty parameter $\beta$, and the baseline parameter $\lambda$.}
    \label{fig:irt-explainer}
\end{figure}

A question $i$ with parameters $z_i$ has an \emph{item characteristic curve} that describes the probability of getting $i$ correct as a function of $\theta$ (essentially, $f$ curried on $z$). \Cref{fig:irt-explainer} shows how the item characteristic curve changes in response to different values of $\alpha, \beta$, and $\lambda$. The latter two parameters are straightforward: a higher difficulty ($\beta$) shifts the curve right, meaning readers need a higher ability to answer correctly. A higher baseline ($\lambda$) shifts the curve upward, meaning the overall probability of answering correctly is increased due to factors like random guessing.

The discrimination parameter $\alpha$ affects the slope of the curve. At $\alpha = 0$, the curve is a flat line. This means that all readers are equally likely to answer correctly regardless of their ability. As $\alpha$ approaches infinity, the curve approaches a vertical line with an x-intercept based on the difficulty. The IRT model of discrimination therefore has a different interpretation than the CTT correlation model: a more discriminative question will more accurately distinguish between readers \emph{of a given level of ability}, as opposed to generally distinguishing between high and low ability readers. We will see the implications of this distinction in the analysis to follow.

\newcommand{\pyirt}{{\small\texttt{py-irt}}}

To fit the IRT model to data, we used the \pyirt{} library\,\cite{lalor-etal-2019-learning}. \pyirt{} uses the Pyro probabilistic programming language\,\cite{pyro2019} to define the IRT model, and it uses Pyro's facilities for stochastic variational inference to train the model. We used \pyirt{}'s default hyperparameters and trained the model for 2,000 epochs. The final output is a set of question parameters $z_i$ for every question in the \trpl{} dataset and a set of ability scores $\theta_j$ for every reader.

\subsubsection{Analysis}

CTT discrimination ($r$) and IRT discrimination ($\alpha$) were least correlated at the extremes of the difficulty distribution. Questions in the middle decile of difficulty had a correlation between $r$ and $\alpha$ of $\data{0.90}$, while questions in the $1^\text{st}$ and $10^\text{th}$ deciles had a correlation of $\data{0.60}$ and $\data{0.01}$, respectively. Put another way, $\alpha$ provides the most distinct signal compared to $r$ at the highest and lowest difficulties.
To explore this phenomenon, we performed another thematic analysis focused on the question: at the tail ends of the difficulty distribution, what characteristics distinguish low-$\alpha$ questions from high-$\alpha$ questions? Specifically, we qualitatively analyzed all questions in the top and bottom deciles of difficulty, sorted by $\alpha$.

In our analysis, the clearest distinction between low and high $\alpha$ questions was the quality of the distractors for multiple-choice questions. (This is a ``soft'' interpretation that is difficult to put into numbers, but you also can review the same data yourself in the artifact accompanying the paper.) By ``quality,'' we mean that good distractors represented plausible misconceptions about Rust, while bad distractors represented either implausible misconceptions about Rust or plausible beliefs about Rust that could arguably be valid conceptions.

\begin{figure}        
    \begin{subfigure}[t]{0.45\textwidth}
        \centering
        \begin{question}
            \questiontype{Multiple Select}
            \begin{questionprompt}
Which of the following are valid reasons for implementing a macro as a
procedural macro instead of a declarative macro?
            \end{questionprompt}
                \begin{qparachoices}
                \rightanswer You want to integrate with Rust's derive system
                \rightanswer Your macro requires nontrivial analysis of the macro user's syntax
                \wronganswer You want to generate variable-length sequences of code
                \wronganswer Your macro requires an entire item as input, not just an expression
                \end{qparachoices}
        \end{question}

        \vspace{0.5em}
        \includegraphics[width=1.5in]{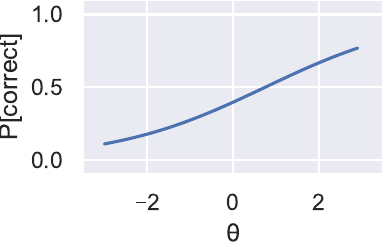}
        \caption{A low-discrimination difficult question with $\alpha = 0.55$}
        \label{fig:macro-question}
    \end{subfigure}
    \hfill
    \begin{subfigure}[t]{0.52\textwidth}
        \centering
        \begin{question}
            \questiontype{Multiple Select}
            \begin{questionprompt}
Say we have a function that moves a box, like this:

\begin{qminted}{rust}
fn move_a_box(b: Box<i32>) {
  // This space intentionally left blank
}
\end{qminted}

\noindent Below are four snippets which are rejected by the Rust compiler. 
Imagine that Rust instead allowed these snippets to compile and run. 
Select each snippet that would cause undefined behavior, or select 
"None of the above" if none of these snippets would cause undefined behavior.
            \end{questionprompt}
             \begin{qchoices}
                \rightanswer\begin{qminted}{rust}
let b = Box::new(0);
let b2 = b;
move_a_box(b);                    
                \end{qminted}
                \rightanswer\begin{qminted}{rust}
let b = Box::new(0);
move_a_box(b);
println!("{}", b);                
                \end{qminted}
                \rightanswer\begin{qminted}{rust}
let b = Box::new(0);
move_a_box(b);
let b2 = b;
                \end{qminted}
                \wronganswer\begin{qminted}{rust}
let b = Box::new(0);
let b2 = b;
println!("{}", b);
move_a_box(b2);
                \end{qminted}
                \wronganswer None of the above
            \end{qchoices}                     
        \end{question}  

        \vspace{0.5em}
        \includegraphics[width=1.5in]{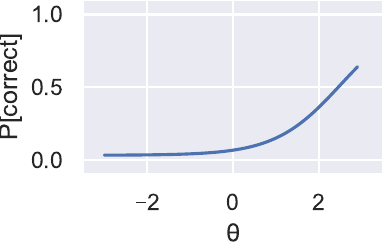}
        \caption{A high-discrimination difficult question with $\alpha = 1.31$}
        \label{fig:box-question}
    \end{subfigure}
    \subfigtocapspace{}
    \caption{Low and high discrimination questions in the highest decile of difficulty, as measured by $\alpha$. Above is the question text, below is the item characteristic curve.}
    \label{fig:discim-comparison}
\end{figure}

For instance, compare the low and high discrimination questions in \Cref{fig:discim-comparison}. The most common incorrect answer to \Cref{fig:macro-question} was to include ``Your macro requires an entire item as input'' as a reason to use a procedural macro over a declarative macro. We considered this incorrect in designing the question because declarative macros \emph{can possibly} take items as input, although it is more \emph{common} for procedural macros to be applied to items. The low discrimination of this question could be reflecting the fact that this distractor is not an unreasonable perspective for a Rust user. By contrast, we carefully designed all the distractors in \Cref{fig:box-question} to test different kinds of undefined behavior involving moves and double-frees. The most common incorrect answer is to exclude one of the correct options, suggesting that this question has high discrimination because only the highest ability Rust learners correctly interpret each situation.

IRT also provides a theoretically more accurate model of reader ability ($\theta$) compared to the reader's raw test scores in CTT. A reader who makes trivial mistakes on easy questions but always aces the hard questions would have a higher estimated ability than a reader in the opposite situation with the same raw score. 
For instance, consider the case of reader \rs{59df64e6-08a0-4638-8399-bcdd138b8d71}, or ``Alice'' for short. Alice has $\theta = 1.10$, which is in the $90^\text{th}$ percentile of abilities in the dataset, but Alice has a raw score of $\overline{x} = 0.625$, which is in the $20^\text{th}$ percentile of raw scores. Alice sometimes makes typos, such as answering \verb|[#test]| when the answer is \verb|#[test]|. Alice's most common mistakes are on tracing-type questions, when she tries to give an output for programs that do not compile (reinforcing the takeaway in \Cref{sec:frequentist-question-analysis}). But Alice correctly answered many difficult questions, and hence was calibrated to a high ability.


\begin{takeaways}
    \item Distractors for multiple-choice questions should be carefully designed to elicit common misconceptions, improving question discrimination.
    \item Item response theory can provide a more accurate model of reader ability than raw scores by accounting for the relative difficulty of each question.
\end{takeaways}

\section{Interventions}
\label{sec:interventions}

\begin{figure}
    \centering
    \fcolorbox{gray}{white}{\includegraphics[width=0.9\textwidth]{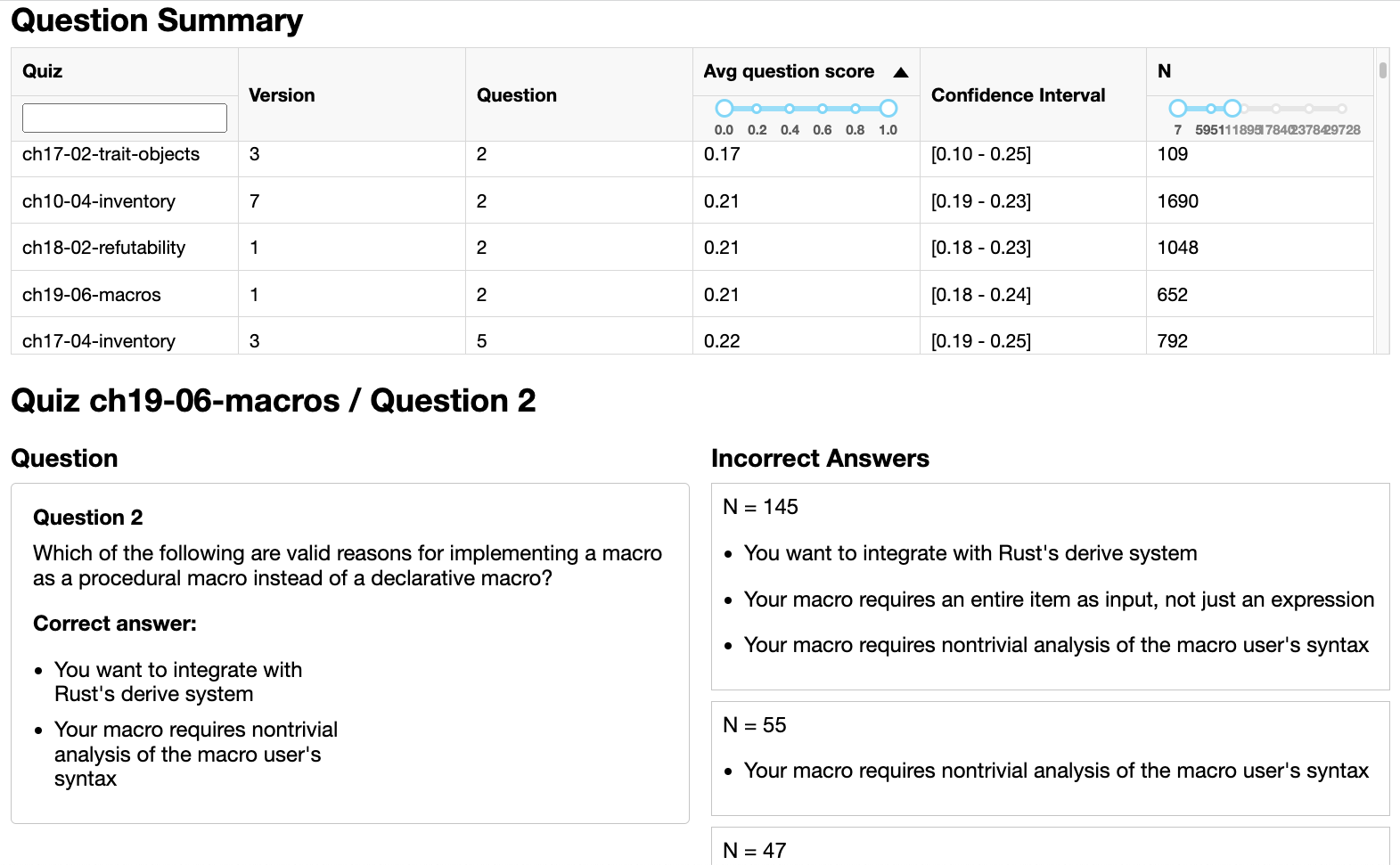}}
    \caption{Screenshot of the data monitoring interface. Quiz-level and question-level statistical summaries are show in tables on the top. Selecting a question brings up the question text and the distribution of incorrect answers. The question in \Cref{fig:macro-question} is selected.}
    \label{fig:question-dashboard}
\end{figure}

\Cref{sec:data-analysis} demonstrates one possible methodology for analyzing a language learning profile. In practice, we imagine that this methodology would exist in an active feedback loop --- authors deploy a text and questions, readers consume the text and answer questions, authors analyze the responses and update the text, and so on. This section describes our attempt to realize this feedback loop in answering RQ3: how can a learning profile be used to improve a PL learning resource?

\subsection{Methodology}

During the deployment of our \trpl{} fork, we monitored the incoming quiz data to identify particularly difficult questions using the dashboard shown in \Cref{fig:question-dashboard}. After identifying a difficult question, we would analyze two key things: (1) the distribution of incorrect responses, and (2) the text surrounding the question in \trpl{}. From this, we would generate a hypothesis about why readers were answering incorrectly, and then encode that hypothesis into a modification to \trpl{}.

\begin{figure}
    \centering
    \begin{subfigure}{\textwidth}
        \centering
        \begin{question}
            \questiontype{Multiple Choice}
            \begin{questionprompt}
                Consider the variables \rss{s2} and \rss{s3} in the following program. These two variables will be located in memory within the stack frame for \rss{main}. Each variable has a size in memory on the stack, \emph{not} including the size of pointed data. Which statement is true about the sizes of \rss{s2} and \rss{s3}? 

                \begin{qminted}{rust}
fn main() {
  let s = String::from("hello");
  let s2: &String = &s;
  let s3: &str = &s[..];
}
                \end{qminted}
            \end{questionprompt}
            \begin{qchoices}[3]
                \rightanswer \rss{s3} has more bytes than \rss{s2}
                \wronganswer \rss{s3} has the same number of bytes as \rss{s2}
                \wronganswer \rss{s3} has fewer bytes than \rss{s2}
            \end{qchoices}
        \end{question}
        \caption{A quiz question with observed high difficulty, only 23\% accuracy pre-intervention.}
        \label{fig:str-layout}
    \end{subfigure}

    \vspace{0.5em}

    \begin{subfigure}[t]{0.47\textwidth}
        \centering
        \includegraphics[width=0.7\textwidth]{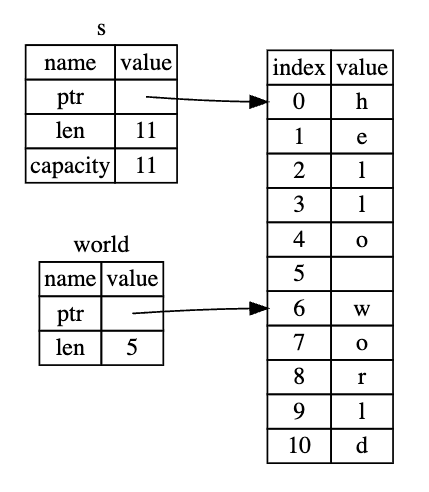}
        \caption{A diagram illustrating the memory layout of string slices in the original \trpl{}.}
        \label{fig:slice-diagram-before}
    \end{subfigure}
    \hfill
    \begin{subfigure}[t]{0.47\textwidth}
        \centering
        \includegraphics[width=\textwidth]{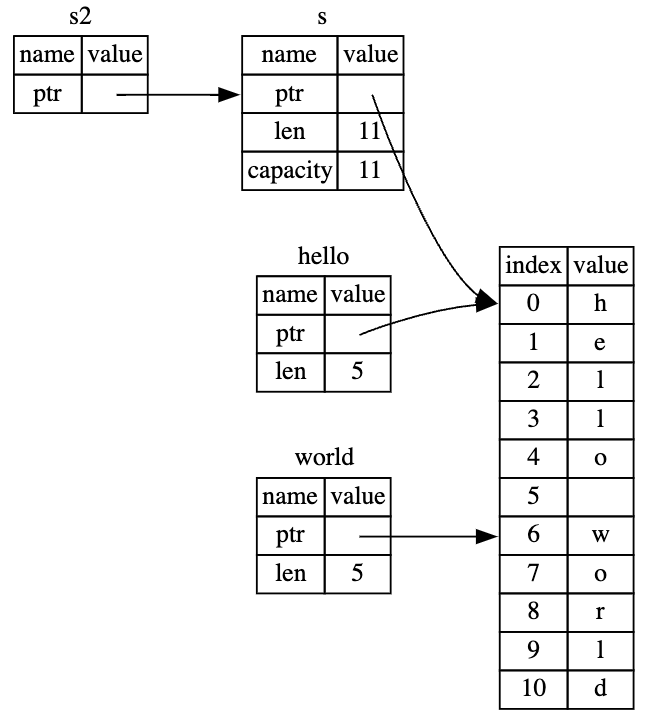}
        \cprotect\caption{Our modified diagram that includes a \verb|&String| reference and a second \verb|&str| reference.}
        \label{fig:slice-diagram-after}
    \end{subfigure}
    \subfigtocapspace{}
    \caption{Anatomy of an intervention: after observing poor performance on the question in \Cref{fig:str-layout}, we modified the diagram in \Cref{fig:slice-diagram-before} to become \Cref{fig:slice-diagram-after}. This intervention improved performance on the question by \data{20\%} ($d = \data{0.41}$).}
    \label{fig:example-intervention}
\end{figure}

\Cref{fig:example-intervention} shows one such intervention. We observed \data{23\%} accuracy on the question in \Cref{fig:str-layout} about the memory layout of references and slices. The key facts being tested are that (a) an \verb|&String| reference is different from an \verb|&str| reference, and (b) that an \verb|&str| reference is a special ``fat'' pointer that carries the length of the data it points to. We observed that \data{51\%} of readers would incorrectly say \verb|s3| and \verb|s2| have the same number of bytes, and \data{26\%} would say that \verb|s3| has fewer bytes than \verb|s2|. 

After reading through the surrounding text (Section 4.3, ``The Slice Type''), we hypothesized that one factor affecting question performance was the diagram in \Cref{fig:slice-diagram-before}. This diagram visualizes the state of memory containing an owned string \verb|s: String| and a string slice \verb|world: &str|. However, this diagram does not visualize a reference to an owned string \verb|&String|. Our theory was that the two incorrect answers could be explained as follows:
\begin{itemize}
    \item ``\verb|&str| has the same number of bytes as \verb|&String|'': a reader had not internalized that \verb|&str| pointers are special fat pointers. The diagram did not help because it did not visualize both a fat pointer and a non-fat pointer.
    \item ``\verb|&str| has fewer bytes than \verb|&String|'': a reader may recall the diagram and remember ``the \verb|String| in the diagram was larger than the \verb|&str|'', conflating a \verb|String| with an \verb|&String|.
\end{itemize}

\noindent We addressed these hypothesized issues by changing the diagram to \Cref{fig:slice-diagram-after}. To emphasize that \verb|String| is distinct from \verb|&str|, we added a second string slice \verb|hello| that points to the beginning of the string. To emphasize that \verb|&String| is distinct and smaller than \verb|&str|, we added a string reference \verb|s2| that points to \verb|s|.

Evaluating an intervention is straightforward: we compare performance on the question before and after deployment of an intervention, like a kind of temporal A/B test. Our methodology is different from a traditional A/B test where users are randomly assigned to one of two conditions --- in our case, at a given point in time, all \trpl{} readers saw the same textbook. Consistent with our goals in \Cref{sec:design-goals}, this approach minimizes the complexity of infrastructure because we do not need to have persistent accounts for readers (otherwise, a reader might run the risk of seeing a different book on their phone and their computer!). However, the downside is that this methodology makes an i.i.d assumption about the temporal stream of readers to draw valid statistical inferences --- we discuss this issue more in \Cref{sec:threats}.
In the case of \Cref{fig:example-intervention}, the string slice diagram intervention was successful --- after deployment, performance on the \Cref{fig:str-layout} question improved from \data{23\%} to \data{43\%}, which is explained by an \data{11\%} decrease in the ``same number of bytes'' answer and a \data{9\%} decrease in the ``fewer bytes'' answer.

In total, we designed \data{12} interventions. We selected places to intervene by looking at the most difficult questions (those with sub-30\% accuracy), and then analyzing the text surrounding the question to see if we could formulate a hypothesis that explains the poor performance. Each intervention consisted of a relatively small change to a localized portion of \trpl{}, usually adding or editing 1-2 paragraphs. Only two interventions consisted of a new subsection. The interventions were designed in two batches that were deployed on \data{November 29, 2022} and \data{June 2, 2023}, respectively.

\subsection{Results}
\label{sec:intervention-results}

\begin{table}
\caption{Effects of interventions on question accuracy. $p$-values are bolded if they are under the $0.05$ significance threshold. $p$-values are corrected for multiple comparisons with the Benjamini–Hochberg method\,\citeyearpar{yoav1995}.}
\label{tab:intervention-results}
\begin{tabular}{l|rr|rr|r|r|r}
\toprule
\textbf{Description} & \textbf{Before} & $\boldsymbol{N}$ & \textbf{After} & $\boldsymbol{N}$ & $\boldsymbol{\Delta}$ & $\boldsymbol{d}$ &  $\boldsymbol{p}$ \\
\midrule
Semver dependency deduplication & 0.18 & 593 & 0.70 & 543 & 0.52 & 1.24 & \textbf{<0.001} \\
Rust lacks inheritance & 0.29 & 234 & 0.74 & 3511 & 0.45 & 1.03 & \textbf{<0.001} \\
Match expressions and ownership & 0.39 & 522 & 0.74 & 4970 & 0.35 & 0.78 & \textbf{<0.001} \\
Send vs. Sync & 0.25 & 639 & 0.49 & 538 & 0.24 & 0.52 & \textbf{<0.001} \\
String slice diagram & 0.23 & 575 & 0.43 & 7188 & 0.20 & 0.41 & \textbf{<0.001} \\
Heap allocation with strings & 0.13 & 265 & 0.27 & 3636 & 0.14 & 0.32 & \textbf{<0.001} \\
Rules of lifetime inference & 0.26 & 177 & 0.40 & 2887 & 0.14 & 0.29 & \textbf{<0.001} \\
Traits vs. templates & 0.38 & 234 & 0.49 & 3511 & 0.11 & 0.21 & \textbf{0.002} \\
Trait objects and type inference & 0.09 & 654 & 0.18 & 544 & 0.09 & 0.27 & \textbf{<0.001} \\
Refutable patterns & 0.17 & 549 & 0.25 & 499 & 0.08 & 0.21 & \textbf{0.001} \\
Declarative macros take items & 0.19 & 340 & 0.23 & 312 & 0.04 & 0.09 & 0.253 \\
Derefencing vector elements & 0.15 & 311 & 0.18 & 4001 & 0.03 & 0.07 & 0.253 \\
\bottomrule
    \multicolumn{5}{r}{\textbf{Grand average:}} &
    \multicolumn{1}{r}{\textbf{0.20}} &
    \multicolumn{1}{r}{\textbf{0.45}}
\end{tabular}
\end{table}

\noindent \Cref{tab:intervention-results} shows the statistical effects of each intervention. Out of \data{12} interventions, \data{10} had a statistically significant difference (as measured by a two-tailed $t$-test, after adjusting for multiple comparisons) between the distribution of scores on the associated question before and after the intervention. The average increase in scores is \data{+20\%} with an average effect size of Cohen's $d = \data{0.45}$. Three interventions initially did not have a statistically significant effect, which we contextualize below.

\paragraph{``Match expressions and ownership''}
 This failed intervention was caused by a simple mistake --- we placed the intervention \emph{after} the location of the relevant quiz, so readers did not observe the intervention before taking the quiz. A reader alerted us to this issue with the quiz feedback mechanism.  We rearranged the quiz and text, and ultimately the intervention was successful.

\paragraph{``Dereferencing vector elements''}
The relevant question asks readers to trace a program that involves the following lines of code:
\begin{minted}{rust}
let mut v = vec![1, 2, 3];
let mut v2 = Vec::new();
for i in &mut v { v2.push(i); }
*v2[0] = 5;
\end{minted}
When asked what the values of \mintinline{rust}{*v2[0]} and \mintinline{rust}{v[0]} are, many readers will say 5 and 1, even though the correct answer is 5 and 5. We hypothesized that readers needed a better explanation of how references to vector elements worked, so we added a paragraph about it to the section on vectors. However, this intervention had no effect.
Upon seeing this lack of effect, we enabled the ``answer justification'' flag described in \Cref{sec:quiz-tool} to better understand readers' reasoning for the 5 and 1 response. We received a number of justifications in this vein like ``the contents of v are copied into v2'' and ``I think v2 is a deep copy of v.''

Based on these responses, we realized that many readers were inferring the type of \rs{v2} to be \mintinline{rust}|Vec<i32>| (a vector of integers) instead of \mintinline{rust}|Vec<&mut i32>| (a vector of mutable references to integers). The goal of the question is not for readers to mentally perform type inference, so in response we modified the question to include type annotations for \rs{v} and \rs{v2}. After modifying the question, scores increased by \data{+9\%} ($p < 0.001$). We do not count this subsequent change as an intervention in \Cref{tab:intervention-results} because the question itself was changed. Nonetheless, this question is a good example of how to use a question's profile to recognize and fix an issue in its presentation.

\paragraph{``Declarative macros take items''}
This intervention relates to the question in \Cref{fig:macro-question} about declarative versus procedural macros. We added a sentence emphasizing the fact that declarative macros can take items as input to the section on macros, but this had no statistically significant effect on the question. We are still in the process of investigating why this intervention was unsuccessful.

\subsection{Analysis}

We provided one example of successful interventions (``string slice diagram''), but this raises the broader question: what are the properties of a successful intervention? Our sample size of 10 successful interventions is relatively small so we cannot yet draw confidently generalizable conclusions, but we can make at least a few preliminary observations.

The two most effective interventions (``semver dependency deduplication'', ``Rust lacks inheritance'') were cases where we realized the question addressed a concept that was not explicitly brought up in the relevant section. The section on Cargo never explicitly said how Cargo uses semantic versioning, and the section on traits never explicitly said that traits do not support specialization of any kind. This fact was revealed by poor performance on questions that tested whether readers would correctly infer these facts from the baseline text, which apparently they would not. Both questions were addressed by adding one paragraph that explains the relevant concept.

\begin{figure}
    \centering
    \begin{subfigure}{0.42\textwidth}
        \begin{framed}
        \begin{footnotesize}
        \begin{singlespace}
        Second, we can see in the signature that add takes ownership of \mintinline{rust}|self|, because \mintinline{rust}|self| does not have an \mintinline{rust}|&|. This means \mintinline{rust}|s1| in Listing 8-18 will be moved into the add call and will no longer be valid after that. So although \mintinline{rust}|let s3 = s1 + &s2;| looks like it will copy both strings and create a new one, this statement actually takes ownership of \mintinline{rust}|s1|, appends a copy of the contents of \mintinline{rust}|s2|, and then returns ownership of the result. In other words, it looks like it’s making a lot of copies but isn’t; the implementation is more efficient than copying.
        \end{singlespace}
        \end{footnotesize}
        \end{framed}
        \caption{Explanation of allocations with string addition in baseline \trpl{}.}
        \label{fig:string-alloc-before}
    \end{subfigure}
    \hfill
    \begin{subfigure}{0.56\textwidth}
        \begin{framed}
        \begin{footnotesize}
        \begin{singlespace}
        Second, we can see in the signature that add takes ownership of \mintinline{rust}|self|, because \mintinline{rust}|self| does not have an \mintinline{rust}|&|. This means \mintinline{rust}|s1| in Listing 8-18 will be moved into the add call and will no longer be valid after that. So although \mintinline{rust}|let s3 = s1 + &s2;| looks like it will copy both strings and create a new one, this statement instead does the following:
        \begin{enumerate}[leftmargin=*,itemsep=0pt]
            \item add takes ownership of  \mintinline{rust}|s1|,
            \item it appends a copy of the contents of  \mintinline{rust}|s2| to \mintinline{rust}|s1|,
            \item and then it returns back ownership of  \mintinline{rust}|s1|.
        \end{enumerate}
        If \mintinline{rust}|s1| has enough capacity for \mintinline{rust}|s2|, then no memory allocations occur. However, if \mintinline{rust}|s1| does not have enough capacity for \mintinline{rust}|s2|, then \mintinline{rust}|s1| will internally make a larger memory allocation to fit both strings.
        \end{singlespace}
        \end{footnotesize}
        \end{framed}
        \caption{Explanation of allocations with string addition after our intervention.}
        \label{fig:string-alloc-after}
    \end{subfigure}
    \subfigtocapspace{}
    \caption{Example of the ``heap allocation with strings'' intervention.}
    \label{fig:string-alloc-intervention}
\end{figure}

Most of the other successful interventions, including ``string slice diagram,'' could best be described as clarifying or enhancing existing content. For instance, the ``heap allocation with strings'' question asks readers to determine the worst-case number of heap allocations that could occur in a program involving string addition. Readers were consistently underpredicting the number, which we hypothesized was caused by the explanation in \Cref{fig:string-alloc-before} which does not make clear that string appending could allocate. The edited explanation in \Cref{fig:string-alloc-after} improved performance by \data{+14\%}. 

One should be cautious in interpreting these interventions as categorically making \trpl{} ``better.'' Most of these interventions made the book longer, and readers' attention is in short supply as we saw in \Cref{fig:last-chapter} --- the statistics indicate whether an intervention helped with a particular learning outcome, but the author must ultimately decide if an intervention is worth the investment. We discuss additional caveats in \Cref{sec:internal-validity}.

\begin{takeaways}
    \item Quiz questions can provide a useful profiling framework for iteratively identifying areas of improvement and then evaluating changes.
    \item A key factor of a successful intervention is drawing the reader's attention to the right concept at the right time in their learning process.
\end{takeaways}

\section{Small-N Simulations}
\label{sec:simulations}

Our goal is to develop a methodology for profiling language learning that is not just applicable to Rust. Clearly, the concept of asking quiz questions can be generalized --- the question templates described in \Cref{sec:quiz-questions} would be valid for the syntax and semantics of most languages. The less obvious aspect of generality is sample size. Rust in 2023 is a popular language, so within a single year, we had tens of thousands of people participate in our experiment. That raises the question in RQ4: how applicable is the question profiling methodology to languages with smaller user bases?

To answer this question, we will revisit a key statistical inference that was part of answering each of the last three RQs. We will use a combination of random sampling and power analysis to simulate how the statistical inference would change at smaller $N$.

\subsection{Simulating Reader Analysis}

For RQ1, we consider \Cref{fig:last-chapter} which visualized the proportion of triers who left the book at a given chapter, showing that Chapters 3 and 4 served as a drop-off point for triers. We will explore: with fewer readers, how would our estimation of the chapter-level drop-off rates change? To answer this question, we will use the same kind of simulation for both RQ1 and RQ2:

\begin{wrapfigure}{r}{0.4\textwidth}
\vspace{-2em}
\begin{minipage}{0.4\textwidth} 
\begin{algorithm}[H]
\caption{Estimating metric error}
\begin{algorithmic}
\State $x \gets f(S)$
\State $r \gets \mathsf{ranks}(x)$ 
\For{$k \in \{10, 100, ..., |S|\}$}
    \For{$i \in {0, 1, ..., 1000}$}
        \Repeat
            \State $S' \gets$ sample($S$, $k$)
        \Until{$S'$ is valid for $f$}

        \State $x' \gets f(S')$
        \State $r' \gets \mathsf{ranks}(x')$
        \State store $|x - x'|$ and $|r - r'|$
    \EndFor
\EndFor
\end{algorithmic}
\end{algorithm}
\end{minipage}
\end{wrapfigure}

Let $S$ be the set of all triers, and let $f : \mathcal{P}(S) \rightarrow \mathbb{R}^n$ be a function that computes a vector of metrics for some $n$, such as $n = $ ``the number of chapters in \trpl{}'' and $f = $ ``the fraction of readers who drop off at each chapter''. Let $\mathsf{ranks} : \mathbb{R}^n \rightarrow \mathbb{N}^n$ return the rank-order of each element in the input. Then our simulation follows the pseudocode on the right.

For several sample sizes $k$, we run 1,000 simulations that compare the metric $f$ on a $k$-sized random sample against the same metric on the full dataset. Each $f$ has a validity condition on the input subset (e.g., every chapter must appear at least once in the sample), so we repeatedly sample until the validity condition is met (i.e., a Las Vegas algorithm). Then we compute the metric error on the sample in terms of both difference in the raw value of the metric ($x$) and different in terms of the rank order of the $n$ items ($r$). The idea is that both raw error and rank error may be relevant depending on the analysis task --- for instance, an author trying to analyze a specific question would care about the raw error, while an author trying to find the hardest question would care about the rank error.

Concretely for RQ1, $n$ is the number of chapters in \trpl{} and $f$ is the fraction of triers who drop-out at a given chapter. The validity condition is that every chapter must appear at least once in the sample. We draw $k$ from a logarithmic range from $100$ to $|S| = \data{32,159}$ (we exclude $k < 100$ for RQ1 because there are very few valid subsets of that size).

\begin{figure}
    \centering
    \begin{subfigure}{0.47\textwidth}
        \includegraphics[width=\textwidth]{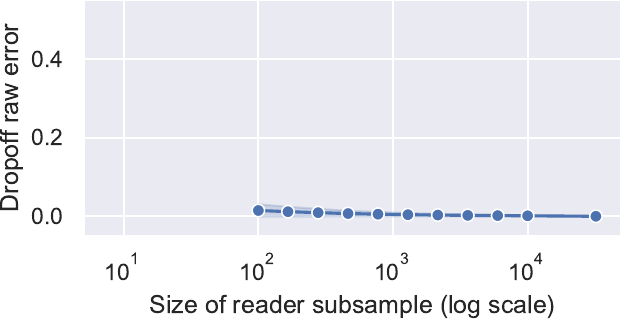}
        \caption{Error in the raw drop-off of each chapter.}
        \label{fig:sim-chapter-raw}
    \end{subfigure}
    \hfill
    \begin{subfigure}{0.47\textwidth}
        \includegraphics[width=\textwidth]{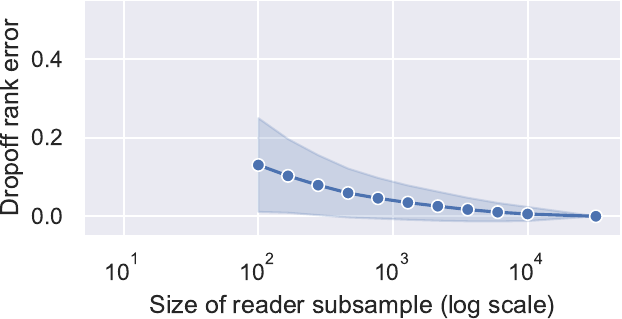}
        \caption{Error in the drop-off rank of each chapter.}
        \label{fig:sim-chapter-rank}
    \end{subfigure}
    \subfigtocapspace{}
    \caption{Error in the simulated estimates of the per-chapter drop-out rates used in \Cref{fig:last-chapter} for RQ1. The shaded region indicates one standard deviation away from the estimated mean at each sample size. The rightmost point in each graph represents the full dataset, and hence has zero error.}
    \label{fig:sim-chapter}
\end{figure}

\Cref{fig:sim-chapter} shows the results of the simulation for RQ1. Both plots show the estimated error for a given $k$ at each dot, as well as $\pm 1$ standard deviation of the estimate in the shaded region.
The raw error is relatively low, being an estimated \data{0.02} at $K = 100$. However, because many chapters have a relatively similar drop-off rate (as shown in \Cref{fig:last-chapter}), the rank error is much larger at small $k$, being an estimated \data{0.13} with standard deviation \data{0.12}.

Our interpretation is that if an author wants to identify individual chapters with an high absolute drop-out rate, then this result suggests that the author needs relatively few readers to do so. If an author wants to understand the specific rank order of chapters by drop-out rate, then this method does not generalize to small $N$.

\subsection{Simulating Question Analysis}

For RQ2, we repeat a similar analysis as for RQ1 except where $n$ is the number of questions in Chapters 3 and 4, and $f$ is the CTT difficulty and discrimination of each question. (We limit our focus to Chapters 3 and 4 because the majority of responses are to those questions, so we can more reliably estimate the error of small-N simulations.) For difficulty, a subset $S'$ is valid if there exists at least one answer for all questions by a reader in $S'$, and we start with $k = 10$. For discrimination, a subset $S'$ is valid if it is possible to compute $r$ for all questions (i.e. there are at least two answers for a given question, and the answers are not all 0 or 1), and we start with $k = 100$.

\begin{figure}
    \centering
    \begin{subfigure}{0.47\textwidth}
        \includegraphics[width=\textwidth]{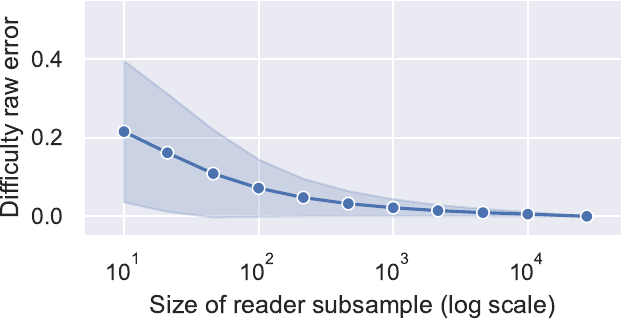}
        \caption{Error in the raw difficulty of each question.}
        \label{fig:sim-difficulty-raw}
    \end{subfigure}
    \hfill
    \begin{subfigure}{0.47\textwidth}
        \includegraphics[width=\textwidth]{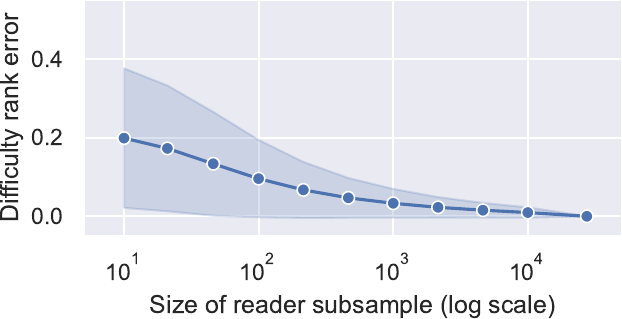}
        \caption{Error in the difficulty rank of each question.}
        \label{fig:sim-difficulty-rank}
    \end{subfigure}

    \vspace{1em}

    \begin{subfigure}{0.47\textwidth}
        \includegraphics[width=\textwidth]{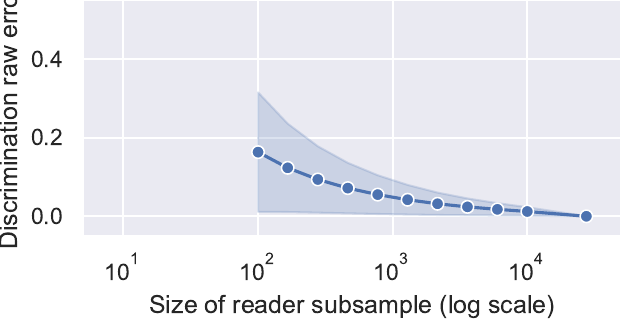}
        \caption{Error in the raw discrimination of each question.}
        \label{fig:sim-discrim-raw}
    \end{subfigure}
    \hfill
    \begin{subfigure}{0.47\textwidth}
        \includegraphics[width=\textwidth]{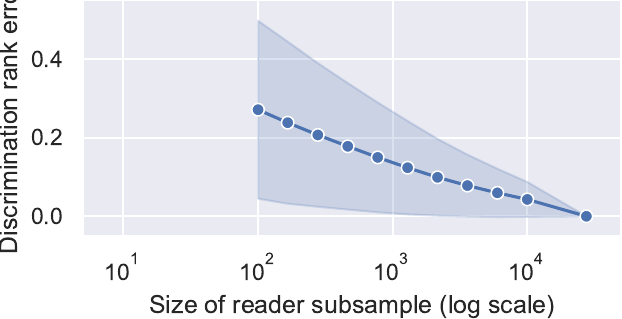}
        \caption{Error in the discrimination rank of each question.}
        \label{fig:sim-discrim-rank}
    \end{subfigure}
    \subfigtocapspace{}
    \caption{Error in the simulated estimates of per-question CTT difficulty and discrimination statistics used in \Cref{fig:question-average-scores}  and \Cref{fig:question-correlation} for RQ2.}
    \label{fig:sim-questions}
\end{figure}

\Cref{fig:sim-questions} shows the results of the simulation for difficulty (top) and discrimination (bottom). At $k = 10$, difficulty has high error in both raw metric (estimated \data{0.22}) and rank (estimated \data{0.20}). At $k = 100$, the error is reduced to an average of \data{0.07}, and further reduced at $k = 1000$ to \data{0.02}. Discrimination has higher error (both raw and rank) at all $k$ compared to difficulty. Even at $k = 1000$, the average discrimination raw error is \data{0.08} and rank error is \data{0.18}.

Our interpretation is that estimates of question difficulty are likely useful for both relative and absolute comparisons at around $k = 100$, but only precise at $k = 1000$. Estimates of question discrimination seem to require at least an order of magnitude more data to be sufficiently precise.

\subsection{Simulating Interventions}

For RQ3, we explore the question: for a statistically significant intervention, what is the smallest number of readers needed to determine that the intervention is statistically significant? This question is a straightforward application of \emph{power analysis}, where one estimates the sample size needed to demonstrate an effect for a given effect size. For a common statistical test like the $t$-test, there is a closed-form function that takes as input an effect size $d$, a significance level $\alpha$, and a test power $\beta$ (commonly $\beta = 0.8$, which we use here). The power analysis returns a sample size $N$ required to identify the effect $d$ at the given significance level $\alpha$ with at least probability $\beta$.

\begin{figure}
\centering
    \begin{minipage}[t]{.47\textwidth}
        \centering
        \includegraphics[width=0.85\textwidth]{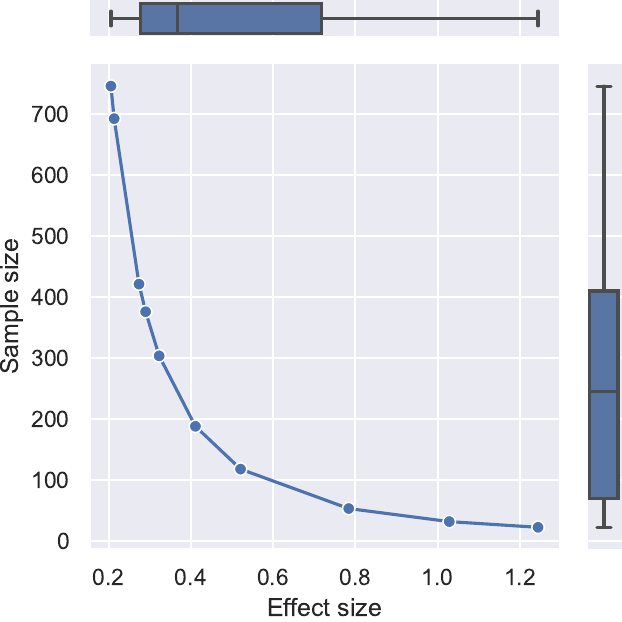}
        \captionof{figure}{Power analysis of the \data{10} statistically significant interventions in \Cref{tab:intervention-results}. Each point is the sample size required to detect (with statistical significance, at an 80\%) probability) an effect with the given effect size.}
        \label{fig:sim-intervention-power-analysis}
    \end{minipage}%
    \hfill
    \begin{minipage}[t]{.47\textwidth}
        \centering
        \includegraphics[width=\textwidth]{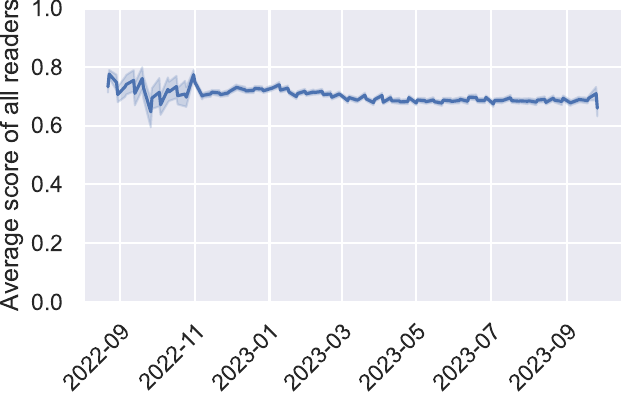}        
        \caption{Average score of all readers by week from the start to the end of the experiment. The shaded region represents a 95\% confidence interval on the estimated average for a given week.}
        \label{fig:scores-over-time}
    \end{minipage}
\end{figure}

We apply the $t$-test power analysis to the 10 statistically significant interventions in \Cref{tab:intervention-results} based on their computed effect sizes. \Cref{fig:sim-intervention-power-analysis} plots the effect size on the $x$-axis against the number of samples required on the $y$-axis, including a marginal distribution of both variables on the top and right. The median required sample size for our interventions is \data{246}. For effect sizes $d > 0.4$, our interventions would have required at most \data{200} readers (that is, 100 before and 100 after the intervention) to demonstrate statistical significance (with 80\% probability). For effect sizes $d < 0.4$, our interventions needed between \data{303} and \data{745} readers to demonstrate statistical significance.

We interpret these results as promising for generality to small $N$ --- an effective intervention (i.e., one with a large effect) does not need thousands of readers, and authors would presumably be trying to make effective interventions. It is true that more data would be needed to measure subtle effects, but we expect that type of effect to be more the domain of researchers than practitioners.

\begin{takeaways}
    \item Estimating chapter drop-out rates and question difficulty can be done with low error at $k \approx 100$, while estimating question discrimination requires more data.
    \item The more effective an intervention, the fewer readers needed to detect statistical significance --- our most effective interventions needed fewer than 200.
\end{takeaways}

\section{Threats to Validity}
\label{sec:threats}

Next, we will discuss threats to validity of the results. 



\subsection{Internal Validity}
\label{sec:internal-validity}

Across all RQs, a key threat to internal validity is the uncontrolled quizzing environment. We had no control over the population of participants, and we had limited control over the conditions under which a participant took a quiz. This raises a few concerns:

\paragraph{What if different kinds of readers answered questions at different times?} For instance, if all the C++ engineers at a company decided to learn Rust at the same time, that could improve quiz scores during a specific period, which could affect the temporal A/B testing used in RQ3. We believe that this is likely not a serious concern at our scale of data collection --- for instance, \Cref{fig:scores-over-time} shows that the average score over time has been remarkably stable since November 2022, when our advertisement went live in \trpl{}.

\paragraph{What if readers used external resources while answering questions?} A reader's response to a question may not truly reflect their understanding of Rust if they used the Rust compiler, Google, ChatGPT, a friend, or other external resources. We address this threat by isolating the quiz-taking environment, as discussed in \Cref{sec:quiz-tool}.

\paragraph{Are the interventions really causing learning gains by improving the textbook?}
An intuitive reading of the interventions in \Cref{sec:interventions} is something like: ``I changed the text, my changes caused readers to develop a better mental model than before, and therefore my intervention is good.'' But we must be careful in  causally attributing observed learning gains to explanation quality. For instance, the act of focusing a reader's attention on a particular concept may itself be sufficient to improve learning outcomes regardless of the explanation's quality\,\cite{vanlehn2003tutoring}. Put another way, interventions are a form of \emph{teaching to the test}. If an intervention is too tailored to the specific question being targeted, then learners are likely not forming a robust mental model.

We managed the issue of learning-from-attention by making interventions have a close to net-zero change in the book's length, which we strove for in our interventions (but did not always achieve). We managed teaching-to-the-test by ensuring that interventions did not change the textbook to trivialize the problems under question, e.g., by adding the answer verbatim to the book.

\paragraph{What if some parts of the experiment interfered with other parts?}
Our experiment was essentially many small experiments running simultaneously. For instance, several interventions were deployed at the same time, and one intervention could theoretically affect other interventions. 
On one hand, this reflects how we expect a learning resource author to use our methodology in practice --- no one would make a single small change and wait a year until making the next change. On the other hand, we cannot easily disentangle the complex web of influence.

It is important to note that all of the interventions described in this paper were relatively small, but the experiment described by \citet{crichton2023aquascope} was running concurrently in the same \trpl{} fork as ours. Crichton et al. deployed a much larger intervention that replaced the entire ownership chapter (Chapter 4) with an alternative based on a new conceptual model of ownership types. This intervention did affect specific questions on ownership (as documented by Crichton et al.), but it did not affect overall question accuracy (as shown in \Cref{fig:scores-over-time}).

\subsection{Construct Validity}
\label{sec:construct-validity}

Several of the statistics in this paper have a relatively uncomplicated relationship to the construct they represent, such as question difficulty (either CTT or IRT) and how hard a reader would find a given question. But there are two constructs in particular worth discussing: 

\paragraph{Are question scores a good proxy for language understanding?}
A fundamental assumption in this paper is that the quiz questions provide a profile of language learning, meaning a reader's answer to a quiz question reflects some aspect of the mental model they are constructing of a language. Of course, a bad question does not test anything useful, and hence we devote \Cref{sec:frequentist-question-analysis} and \Cref{sec:bayesian-question-analysis} to the analysis of question quality. 
But more fundamentally, our format of quiz questions is limited to problems that are small, self-contained, and answerable in under a minute. Deep expertise in a programming language is most evident in the design of complex systems over long time-scales. Quiz questions test for small slices of theoretical understanding, as opposed to the practical knowledge of working with a tool every day. In that sense, we see quiz questions as complementary to other sources for language usability evaluation.

\paragraph{Are the discrimination metrics a good proxy for question quality?}
While discrimination is commonly used in psychometrics for item analysis, it is not a flawless metric of quality. Discrimination metrics reveal correlation more than causation, and correlation can easily miss unmeasured causal factors. For instance, a question may be highly discriminative because it tests whether a learner comes from a particular background, as opposed to whether the learner understands a textbook well\,\cite{masters1988discrim}. To that end, we tried to provide a balanced look at discrimination in this paper by considering two different models. Discrimination should be best understood a useful signal for filtering data, but not an ultimate objective to be fully optimized.


\subsection{External Validity}

The main threat to external validity is whether these methods scale to languages with smaller user bases. We analyze and discuss this threat extensively in \Cref{sec:simulations}.

\section{Related Work}
\label{sec:relatedwork}

Methodologically, our experiment is most related to computing education research on e-books and item analysis. Philosophically, our experiment is most aligned with human factors research on the adoption of programming languages. We focus on the former here, and save the latter for \Cref{sec:discussion}.

Many educators and researchers have developed e-books for teaching both computer science generally and programming languages specifically, usually targeting the introductory level. Most similar to our work is Runestone\,\cite{ericson2020runestone}, an e-book platform used to teach various CS courses with support for interactive content such as quizzes. The research performed with Runestone is complementary to ours --- for instance, \citet{ericson2022ebook} found that doing well on multiple-choice questions in the e-book was strongly correlated with midterm performance. Our research can hopefully inform the development of content on platforms like Runestone. Our results also provide additional insight for the deployment of interactive e-books ``in the wild,'' as opposed to school contexts where students are required to read to the end of the book.

CS education researchers frequently use item analysis, and often specifically item response theory, to analyze (or ``validate'') tests of programming knowledge\,\cite{porter2019validate,kong2022validate}. For instance, \cite{xie2019irt} use IRT discrimination to identify and eliminate problematic questions on a test of introductory CS knowledge. We extend this line of work with data from a much larger scale on programming problems aimed at a more experienced audience of programmers.

Our goal is to treat Rust only as an exemplar of language learning writ large, but our results nonetheless are complementary to the growing body of research on the usability of Rust. Researchers have used surveys\,\cite{fulton2021}, social media\,\cite{zeng2018}, and StackOverflow\,\cite{zhu2022} to analyze Rust's learning curve, similarly finding that ownership is a frequent challenge for learners. \citet{crichton2023aquascope} demonstrated one example of a successful intervention in \trpl{} limited to ownership. Our work goes beyond it in several ways. First, we analyze \trpl{} at a much broader scope than a single chapter, enabling us to examine reader drop-off and question discrimination. Second, we expand beyond classical test theory to incorporate the nuanced IRT model of readers and questions. Most of all, our hope is to design a methodology that can work for languages that do not enjoy Rust's scale; \Cref{sec:simulations} shows that we appear to succeed.




\section{Discussion}
\label{sec:discussion}

A decade ago, \citet{meyerovich2013adopt} demonstrated the value of using surveys to understand programmers' attitudes towards programming language adoption, arguing that ``these methods will be increasingly valuable going forward, especially given the popularity of the Internet and online courses.'' History has supported their argument --- as we cited in the introduction, one can now find survey data for dozens of programming language communities.

We believe that programming language learning is in need of an analogous set of techniques for peeking into the minds of programmers. 
%
%
In this paper, we propose the method of \emph{profiling} language learners via interactive quizzes embedded in learning resources. We analyzed over 1,000,000 quiz responses from a year of data collection from \textit{The Rust Programming Language}. We identified several interesting aspects of Rust language learning, such as where readers drop-out of the book and which kinds of questions were most and least discriminative of Rust ability. We showed how to use this quiz data in an iterative loop to improve learning outcomes with targeted interventions.

This method is \emph{simple}: the infrastructure is easy to setup, and all one needs to do is write some quiz questions and pay for a telemetry server. This method is \emph{engaging}: many readers voluntarily opt to use text with quizzes when given the choice. Readers only spend a median of \data{29 seconds} per question. This method is \emph{generalizable}: the question format works with any language, and most of the statistical inferences only need a few hundred readers to be accurate. We encourage authors of language learning resources to try adding quizzes to your book; if you already use mdBook, then you can try out our free and open-source quiz plugin: \url{https://github.com/cognitive-engineering-lab/mdbook-quiz}

The goal of this paper is to provide a simple and replicable blueprint, but we have left many unanswered questions about how people learn programming languages. Avenues for future work include:
    
\paragraph{How can quiz data inform PL design?} In this work we treat the language (Rust) as an artifact with fixed rules that must be conveyed as-is to learners. Incorrect answers are treated strictly as misconceptions. But for language designers, an incorrect answer could also be a statement of preference: \textit{I thought the language should work this way, but it doesn't seem to.} These types of insights can influence language design --- for instance, the lifetime analysis within Rust was made flow-sensitive for this reason. According to \citet{nllrfc}:
\begin{quote}
``Part of the reason that Rust currently uses lexical scopes to determine lifetimes is that it was thought that they would be simpler for users to reason about. Time and experience have not borne this hypothesis out: for many users, the fact that borrows are ``artificially'' extended to the end of the block is more surprising than not.''
\end{quote}

\noindent Future work should investigate how profiling learning can meaningfully inform PL design. However, this process will inevitably require good judgment, as programmers are not necessarily consistent in their beliefs about language design\,\cite{tunnell2017crowdsource}.

\paragraph{Does adding quizzes to PL textbooks facilitate learning \textit{per se}?} Previous studies have reported that interspersing questions into educational materials has improved learning outcomes\,\cite{rothkopf1967questions,andre1979questions}. Intuitively, answering a question will prompt a learner to engage further with the material covered by the question, and we should expect PL quiz questions to be no different. But the specific learning outcomes are contingent on the types of questions asked, and future work ought to investigate what types of questions are most beneficial to learning. (As \Cref{sec:ctt-discrimination} shows, we can likely rule out ``does-it-compile'' questions.)

\paragraph{What is the right balance of effort to payoff for PL quizzes?} Quiz questions likely give diminishing returns to both the learner and educator as they increase in either quantity or difficulty. Unlike a classroom setting, a PL textbook cannot mandate its readers to answer a set of questions. It is essential therefore to find the sweet spot that maximizes both learning outcomes (for the learner) and insight into the learning process (for the educator).

\paragraph{Why do learners drop out of a PL textbook?}
All improvements in learning outcomes are for naught if a learner ultimately gives up on a language. It would be valuable to understand the specific factors that cause learners to stop reading a PL textbook. Is it the language's perceived difficulty or complexity? Perhaps a perceived lack of utility or relevance to the learner? Or perhaps the learner is not giving up at all, and simply feels they have learned enough to attempt their current task. Future work can consider some form of lightweight ``exit survey'' to capture these attitudes.


\begin{acks}
This work was partially supported by a gift from Amazon and by the US NSF under Grant No.~2319014. We are deeply grateful to the many contributors to \emph{The Rust Programming Language}, especially Carol Nichols who allowed us to advertise in the main digital edition. We are also grateful to the 62,526 anonymous Rust learners who participated in the experiment --- this work is ultimately for them.
\end{acks}

\bibliography{bib}

\newpage

\appendix
\section{Appendix}

\subsection{Highest Correlating Subset}
\label{sec:highest-correlating-subset}

These six questions were the subset with the highest correlation ($r = 0.91$) between reader's average scores on the subset and reader's average scores overall. \\

\noindent\begin{question}
    \questiontype{Multiple Choice}
    \begin{questionprompt}
        Consider these two methods that increment a field of a struct. Which style would be more idiomatic for Rust?

\begin{qminted}{rust}
struct Point(i32, i32);
impl Point {
  fn incr_v1(mut self)  { self.0 += 1; }
  fn incr_v2(&mut self) { self.0 += 1; }
}
\end{qminted}
    \end{questionprompt}
    \begin{qchoices}
\rightanswer \rss{incr\_v2}
\wronganswer \rss{incr\_v1}
\wronganswer Both are idiomatic
\wronganswer Neither are idiomatic        
    \end{qchoices}
\end{question} \\

\noindent\begin{question}
    \questiontype{Tracing}
    \begin{questionprompt}
\begin{qminted}{rust}
fn f(x: i32) -> i32 { x + 1 }
fn main() {
  println!("{}", f({
    let y = 1;
    y + 1
  }));
}
\end{qminted}
    \end{questionprompt}
    \begin{qchoices}
\rightanswer DOES compile with output 3
\wronganswer Does NOT compile
    \end{qchoices}
\end{question} \\

\noindent \begin{question}
    \questiontype{Multiple Select}
    \begin{questionprompt}
        Which of the following are valid reasons for implementing a macro as a
procedural macro instead of a declarative macro?
    \end{questionprompt}
    \begin{qchoices}
\rightanswer You want to integrate with Rust's derive system
\rightanswer Your macro requires nontrivial analysis of the macro user's syntax
\wronganswer You want to generate variable-length sequences of code
\wronganswer Your macro requires an entire item as input, not just an expression 
    \end{qchoices}
\end{question} \\

\noindent\begin{question}
    \questiontype{Multiple Choice}
    \begin{questionprompt}
In some concurrency APIs, a mutex is separate from the data it guards. For example, imagine a hypothetical Mutex API like this:

\begin{qminted}{rust}
let mut data = Vec::new();
let mx: Mutex = Mutex::new();
{
    let _guard = mx.lock();
    data.push(0);
}
\end{qminted}

Which of the following best describes why Rust uses \rss{Mutex<T>} instead of just \rss{Mutex}?
    \end{questionprompt}
    \begin{qchoices}
\rightanswer To prevent accessing a mutex's data without locking the mutex
\wronganswer To require fewer calls to mutex methods
\wronganswer To improve the efficiency of concurrent programs with mutexes
\wronganswer To prevent a mutex's data from being moved between threads    
    \end{qchoices}
\end{question} \\

\noindent\begin{question}
    \questiontype{Multiple Choice}
    \begin{questionprompt}
At the end of this function, Rust will call `drop` on which variables:

\begin{qminted}{rust}
fn main() {
  let s1 = String::from("a");
  let s2 = String::from("b");
  let s3 = s2;
}
\end{qminted}
    \end{questionprompt}
    \begin{qchoices}
\rightanswer \rss{s1} and \rss{s3}
\wronganswer \rss{s1}, \rss{s2}, and \rss{s3}
\wronganswer \rss{s2} and \rss{s3}
\wronganswer \rss{s3}
    \end{qchoices}
\end{question} \\

\noindent\begin{question}
    \questiontype{Short Answer}
    \begin{questionprompt}
What is the name of the command-line tool for managing the version of Rust on your machine?
    \end{questionprompt}
    \begin{qchoices}
\rightanswer rustup
    \end{qchoices}
\end{question}

\end{document}